\documentclass[12pt]{article}
\usepackage{graphicx}% Include figure files
\usepackage{epstopdf}
\usepackage{psfrag}                  % Replace text by LaTeX in EPS figures
\usepackage{url}                        % Deal with URL in an easy manner
\usepackage{times}                    % Use Times font
\usepackage{makeidx}               % allows index generation
\usepackage{multicol}                % used for the two-column index
\usepackage[bottom]{footmisc} % places footnotes at page bottom
\usepackage{fancyhdr}
\RequirePackage{subfigure}
\newcommand{\be}{\begin{equation}}
\newcommand{\ee}{\end{equation}}
\newcommand{\ba}{\begin{eqnarray}}
\newcommand{\ea}{\end{eqnarray}}

\newcommand{\vp}{\varphi}

\newcommand{\al}{\alpha}

\newcommand{\lbd}{\lambda}

\newcommand{\prt}{\partial}

\begin{document}

\begin{center}
{\Large{\bf Endogenous versus Exogenous Origins of Diseases}
\\[5mm]
D. Sornette$^1$, V.I. Yukalov$^{1,2}$, E.P. Yukalova$^3$, J.-Y. Henry$^4$, \\
D. Schwab$^5$, and J.P. Cobb$^6$} \\ [3mm]

{\it
$^1$ ETH Zurich, Department of Management, Technology and Economics \\
CH-8032 Z\"urich, Switzerland\\ [2mm]

$^2$Bogolubov Laboratory of Theoretical Physics, \\
Joint Institute for Nuclear Research, Dubna 141980, Russia \\ [2mm]

$^3$Department of Computational Physics, Laboratory of Information
Technologies, \\
Joint Institute for Nuclear Research, Dubna 141980, Russia \\ [2mm]

$^4$Institut de M\'edecine et Sciences Humaines (IMH SA),\\
10 route de Bremblens, CH-1026 Echandens, Switzerland \\ [2mm]

$^5$ Department of Physics and Astronomy,\\University of California,
Los Angeles, California 90095, USA\\ [2mm]

$^6$ Cellular Injury and Adaptation
Laboratory,\\Washington University in St. Louis, St. Louis, MO, USA}

\end{center}

\vskip 1cm

\today
\vskip 0.5cm

\begin{abstract}
Many illnesses are associated with an alteration of the immune
system homeostasis due to a combination of factors, including
exogenous bacterial insult, endogenous breakdown (e.g., development
of a disease that results in immuno suppression), or an exogenous hit
like surgery that simultaneously alters immune responsiveness and
provides access to bacteria, or genetic disorder. We conjecture that,
as a consequence of the co-evolution of the human immune system
with the ecology of pathogens, the homeostasis of
the immune system requires the influx of pathogens. This allows the immune
system to keep the ever-present pathogens under control and to react
and adjust fast to bursts of infections. We construct the simplest and
most general system of rate equations which describes the dynamics of
five compartments: healthy cells, altered cells, adaptive and innate
immune cells, and pathogens.
We study four regimes obtained with or without  auto-immune disorder and with or without
spontaneous proliferation of infected cells. For each of the four regimes,
the phase space is always characterized by four
(but not necessary identical) coexisting stationary
structurally stable states. Over all four regimes among the $4 \times 4$ possibilities,
there are only seven different states that are naturally described by the model:
(i) strong healthy immune system, (ii) healthy organism with evanescent
immune cells, (iii) chronic infections, (iv) strong infections, (v) cancer, (vi) critically ill state
and (vii) death. Our description provides a natural framework for describing
the relationships and transitions between these seven states.
The analysis of stability conditions
demonstrates that these seven states depend on the balance between the
robustness of the immune system and the influx of pathogens. In particular,
the healthy state A is found to exist only under the  influence of a
sufficiently large pathogen flux, which suggests that
health is not the absence of pathogens, but rather a strong ability to find balance by
counteracting any pathogen attack.

\end{abstract}

\section{Introduction and general background}

Our goal is to develop a model of a biological organism,
defined as the collection of organs, tissues, cells, molecules involved
in the reaction of the body against damaging stressors, which can take
the form of (i) foreign biological material, (ii) damaged, aging and/or
aggressive inner biological material and (iii) inorganic substances.

In a first broad-brush approach, the occurrence of
illness is usually attributed to the following factors, which often act
in combination, sometimes synergistically.
\begin{enumerate}
\item
Microorganisms (bacteries, viruses, fungi, parasites) and the more
recent extensions to prions. This reflects the germ theory of diseases
which states that many diseases are caused by microorganisms, and that
microorganisms grow by reproduction at the expense of the host, rather
than diseases being spontaneously generated. We refer to the microbial
origin of diseases as one of the  ``exogenous'' insults to which the body
is subjected.

\item  Exogenous accumulative load
of stressors in the environment, over-work, over-eating
and other excesses, psychological
and emotional factors (anger, fear, sadness, and so on)
may lead to fatigue and/or epigenetic expressions.
These various stressors impact the immune system by
destabilizing the feedback processes
of the cell-cell communication paths known to exist for the immune system:
direct interaction with neighbor or self (juxtacrine and autocrine, respectively),
short distances (paracrine, such as neurotransmitters), long distance endocrine
(via hormones) and long distance nerves (such as vagus nerve).

\item Genetic variation (which include
polymorphisms that affect outcome, as well as ``disorders'' or
mutations, per se) caused by an unwelcomed mutation as in cancers, by the
accidental duplication of a chromosome, or the defective genes inherited
from the person's parents (hereditary disease).
One should distinguish between two types of hereditary diseases, those
that are immuno-genetic diseases and immune
disorders. The former are the expression of a major immune disfunction
(as for example in trisomy 21 or the Turner syndrome) or of the
deficiency of an enzyme essential to life (such as in mucoviscidosies,
leucodystrophies, Wilson disease, and so on). These affections are
triggered within the first few months of life and, for most of them,
are life-threatening. The only treatment, which is at its beginnings,
consists in gene transfer via a viral vector. In contrast, acquired
immune disorders appear later, often during
adult life. These diseases frequently are expressing defects affecting
those parts coding the immune system in the
region HLA of chromosome 6. At present, more than 45
auto-immune illnesses have been
linked to the idiosynchratic characteristics of this region.
We refer to the genetic disorder origin of diseases as ``structural''
and the present model offers a relevant framework.

\item
Senescence, engineered death (cells ``wear out'').
\end{enumerate}

Most changes that occur in response to stimuli are adaptive,
allowing the system to return back to its ``attractor'' state,
often referred to as homeostasis.  Damaging responses can result
from host failure, overwhelming stimulus, or a combination of the two,
and as a consequence, the host state may then not return to the original
attractor, perhaps a new (different) attractor/state. A normal system
may encounter an overwhelming stimulus (e.g., sepsis). Rarer occurrence
in everyday life are the transient and partial failure of the immune
system which may lead to various degrees of inappropriate response:
(i) different hypersensitivities, in which the system responds
inappropriately to harmless compounds (allergies and intolerances),
(ii) autoimmunity, in which the immune system (mainly via its antibodies)
attacks its own tissues, e.g.,  systemic lupus erythematosus,
rheumatoid arthritis, chronic lymphocytic thyroiditis and myasthenia gravis,
(iii) immunodeficiency, in which parts of the immune system fail to
provide an adequate response (an example of failure to respond is cancer,
in which the immune system fails to recognize the tumoral cells as
dangerous).

Our hypothesis developed below in the mathematical model is based on the
recognition that the human body and immune system function in a form of
homeostasis, a dynamical equilibrium whose balance is continuously
subjected to various external and internal stresses. Genuine epidemic
diseases are relatively rare compared with illnesses that can be
attributed to the transition of the homeostasis to an unbalanced state.
Thus, given that the absence of illness, defined as health, is a form of homeostasis,
living organisms regulate their internal environment so as to
maintain a stable condition, by means of multiple dynamic equilibrium
adjustments controlled by inter-related mechanisms of regulation. Main
examples of homeostasis in mammals include the regulation of the amounts
of water and minerals in the body by osmoregulation happening in the
kidneys, the removal of metabolic waste by excretion performed by
excretory organs such as the kidneys and lungs, the regulation of body
temperature mainly done by the skin, the regulation of blood glucose
level, mainly done by the liver and the insulin secreted by the pancreas.
The mechanism of homeostasis is mostly negative feedback, according to
which a system responds in such a way as to reverse the direction of
change. There are also positive feedback systems, although they are
apparently less frequent (for example, uterine contractions during
parturition). But in contrast with
other homeostatic systems, the immune system is probably better described
by the fact that ``The system never settles
down to a steady-state, but rather, constantly changes with local flare
ups and storms, and with periods of relative quiescence," as expressed
in the mathematical models developed in (Perelson, 2002;
Perelson and Weisbuch, 1997; Nelson and  Perelson, 2002). Our proposed
model sees these  ``flares and storms'' as transient nonlinear adjustments
to fluctuating exogenous fluxes. We delay to a sequel paper the analysis
of the dynamics of our mathematical system under the influence of time varying pathogen
fluxes and under varying conditions. Here, we construct the model based
on general concepts, and classify all its equilibrium states, each of them
associated with a large class of affections.

The paper is organized as follows. Section 2 articulates our endogenous versus
exogenous hypothesis, motivating the construction of the model presented in section 3.
Section 3.1 gives the general mathematical framework in terms
of nonlinear kinetic equations of the concentrations of five different
classes of cells (or biological compartments). Section 3.2 specifies
the kinetic rates of each elementary interaction between these five species
of cells leading to the general form of the equations given in Section 3.3.
Section 3.4 provides their dimensionless reduced form and section 3.5
presents our a priori expectations on the behavior of this model.
Section 4 (respectively 5) presents the properties of the equilibrium states found when
infected cells are not reproducing (respectively are reproducing) by themselves.
Section 6 concludes. The Appendix presents the Lyapunov stability
analysis around the fixed points.

\section{Hypothesis of endogenous versus exogenous origins of diseases}

The human immune system is subjected to incessant ``attacks'' by antigens of
many different forms. Here, we use the term ``antigen'' to refer to
all substances which are recognized as ``non-self'' endowed with an
antigenic functionality, which includes pathogens, cell debris
as well as toxins.
A typical human being carries about $10^{14}$ bacteria (of course many
of them symbiotic) and probably many more viruses. In comparison, the
number of cells constituting the self is only about $10^{13}$ (ten trillions).
In this context, the immune
system is constantly challenged, it is performing a continuing ``fight''
and adaptation to ensure the integrity of the body. This can be viewed
as a continuous flux of ``small'' perturbations to which the immune system has
learned to adapt and to more or less cope with (Mazmanian et al., 2005;
Palmer et al., 2007). Most of the time, an individual is in good health.

A first scenario is that a normal system may encounter
an unusually strong insult leading to a disease.  This is the most
generally accepted scenario. As an example, consider the extreme case
of a healthy human landing in the middle of
a virulent cholera epidemic in Africa or sustaining severe injury in a car crash.
We refer to this situation as an ``exogenous'' shock as the
immune system has suddenly to cope with a serious attack from the
corresponding pathogens or physical destructions.

Let us now consider a second hypothetical scenario, which we call
``endogenous.'' With the same typical fluxes of antigens, by stress or
other destabilization external influences, or simply by chance, sometimes
the immune system of an individual  wanes and he becomes sick. This can occur
after some fatigue (overwork, bad eating, stress, pain, psychological
effects, and so on). In such a case, we cannot say that the illness is
really due to a specific microbial attack, the microbes have been already
present before; it is only that the immune system has gone a bit down. Perhaps,
a succession of small random perturbations may add ``coherently'' in an
unlucky run of random occurrences and lead to sickness, as suggested by
system models of other complex systems (Sornette and Helmstetter, 2003;
Sornette et al., 2003, 2004; Sornette, 2005). A given disease may have
several completely distinct origins, e.g., liver cirrhosis which can
be due to chronic viral infection, alcoholism, eating excesses (NASH,
Non- Alcoholic Steatorrhoeic Hepatitis) or auto-immunity. Often, the
revealed sickness could also be viewed as a positive, i.e., robust
response (via inflammation, fever) to an invading organism. The perception
of being ill and the need for rest may be adaptive and part of the overall
dynamical response towards homeostasis. We refer to this class of events
in which the immune system homeostasis is perturbed away from its
domain of dynamical balance as an endogenously generated illness.

We are aware that this definition of  ``endogenous'' may appear fuzzy,
as a result of the many possible variations. If we have to be blunt, 
endogenous is everything that is not exogenous! Previous works
have found specific dynamical diagnostics of endogenous evolution versus
exogenous shock  (Sornette and Helmstetter, 2003;
Sornette et al., 2003, 2004; Sornette, 2005), which we will not pursue
here as our emphasis is in the classification of the different
homeostatic equilibrium states. We will defer to a future work 
for the dynamical aspects of the model.

We do not address here aging which also leads to failing immunity, as
it is associated with a slow secular non-stationary state whose effect
we neglect for time scales shorter than the average lifetime. Indeed, even the
slow development of chronic diseases found in our model classification occurs
over times scales of no more than a few decades and can thus be (marginally) separated
from aging per se.

Based on these observations,
our hypothesis is that a fundamental understanding of
health, of illnesses, and of the immune system, requires an approach
based on the concept of a self-organization of the immune system into
a homeostasis under the continuous flux of external influences. We
propose that many of the illnesses carried by humans are in significant
part endogenous in nature and that, in order to understand the response
of a human immune system to exogenous shocks, it is necessary to understand its
endogenous organization and the fact that spontaneous
fluctuations in the dynamics of the immune system will occur as a result
of many coherently interfering factors, which may lead to any of
a variety of pathological states mentioned above.
We propose to go beyond the
exogenous-environmental-structural origins of illnesses summarized
in the introductory section, to encompass
a complex system approach in which the human immune system and the whole
body are regulated endogenously under the influence of a continuous
and intermittent flux of perturbations. We conjecture that,
if the regulatory immune system was not constantly subjected to antigens,
it would probably decay in part and the defense would go down as
its adaptive part would not be sustained, thus becoming vulnerable to
future bursts of pathogen fluxes.  Thus, we claim that the correct
point of reference is not to consider a microbe-free or gene-defect-free
body, but a homeostatic immune system within a homeostatic body, under the
impact of many fluxes, in particular fluxes of pathogens and of stresses
taking many of the forms mentioned above. An analogy may serve
to illustrate our point: consider the fate of the bones and muscles
of a typical individual,
which need to be continuously under the influence of a suitable
gravity field, i.e., under stress (in the mechanical sense of the term!).
Otherwise, as demonstrated by astronauts under zero-G,
loss of bone and muscle, cardiovascular deconditioning, loss of red
blood cells and plasma, possible compromise of the immune system,
and finally, an inappropriate  interpretation of otolith system signals
all occur, with no appropriate counter-measures yet known (Young, 1999).

We conjecture
that the healthy individual has a homeostatic immune system working
at a robust level of stability, allowing it (1) to keep ever present
pathogens under control and (2) to react and adjust fast to new
infections and other stresses. Under such fluxes, the
complex regulatory immune systems exhibit spontaneous
fluctuations and shocks, in the form of illnesses, which are themselves
modulated by other factors. In other words, we propose to view
illnesses as emergent properties of a complex interplay and balance
between the immune system, the pathogens and the other stress factors.
We hypothesize that the
emergent properties of the normal system (health) are different from
those of the damaged system (disease). Recent works on other complex
systems suggest that there are ways to find specific diagnostics
distinguishing between endogenous and exogenous causes of crises and
to derive precursors and possible remedies (Sornette and Helmstetter,
2003; Sornette et al., 2003, 2004: Sornette, 2005).
Our hypothesis of an endogenous origin of illnesses resonates with
the recent proposal that the adaptive immune system may have
evolved in vertebrates to recognize and manage the complex
communities of coevolved bacteria (McFall-Ngai, 2007): more than
2000 bacterial species have been found to be typically associated as
partners to a human immune system, compared with fewer than 100
species of identified human bacterial pathogens, with exposure to them
that are rare and transient.

Our endo-exo hypothesis extends the ``hygiene hypothesis''
(Strachan, 1989), which states that modern medicine and sanitation
may give rise to an under-stimulated and subsequently overactive
immune system that is responsible for high incidences of immune-related
ailments such as allergy and autoimmune disease. Strachan (1989) thus
proposed that infections and unhygienic contact might confer
protection against the development of allergic illnesses.
Researchers in the fields of epidemiology, clinical science, and
immunology are now exploring  the
role of overt viral and bacterial infections, the significance of
environmental exposure to microbial compounds, and the effect
of both on underlying responses of the innate and adaptive
immunity (Schaub, 2006). Bollinger et al. (2007) has
recently suggested that the hygiene hypothesis may explain
the increased rate of appendicitis ($\sim 6\%$ incidence) in
industrialized countries, in view of the important immune-related function
of the appendix. Our endo-exo hypothesis is also a distinct generalization
of Blaser and Kirschner (2007)'s proposal that
microbial persistent relationships with human hosts represent a co-evolved
series of nested equilibria, operating simultaneously at
multiple scales, to achieve an overall homeostasis.
Blaser and Kirschner (2007) emphasize the maintenance of persistent
host-adapted infections via ESS (Evolutionary Stable Strategies, a subset
of Nash equilibria in game theory) played by different pathogens between
themselves and the host selected for
human-adapted microbes. In contrast, we emphasize the co-evolution
of the immune system and the pathogens as a key element to ensure
a stable and robust homeostasis.

The next section presents a model which embodies these ideas
in the simplest possible framework, that of kinetic reactions occurring
between five different types of cells. The interactions between these five
compartments are thought to be the outcome of the co-evolution between the
organism, its immune system and the pathogens. However, it is not the
purpose of our model to describe this
evolution, since the time scales modeled here are typical of the recovery times
following pathogen insults, which are much shorter than those involved in evolution.

As reviewed by Louzoun (2007),
mathematical models used in immunology and
their scope have changed drastically in the past 10 years.  With the
advent of high-throughput methods, genomic data, and explosive
computing power, immunological modeling now uses high-dimensional
computational models with many (hundreds or thousands of coupled
ordinary differential equations (ODEs))
or Monte Carlo simulations of molecular-based approaches.
Here, in contrast, we develop a five-dimensional system of the
coupled normal cells-immune system(s)-infected cells-pathogens, which
is a direct descendent of the classical models that
were based on simple ODEs, difference equations,
and cellular automata. While the classical models
focused on the simpler dynamics
obtained between a very small number (typically two or three) of
reagent types (e.g. one type of
receptor and one type of antigen or two T-cell populations) (Louzoun, 2007),
we are more ambitious with the goal of framing an holistic approach
to the homeostasis of the immune system(s) seen to be continuously interacting
with other compartments of the organism and with pathogens.
Our motto is well-captured by the quote from the biocyberneticist
Ludwig von Bertalanffy: ``Over-simplifications, progressively corrected
in subsequent development, are the most potent or indeed the only means
toward conceptual mastery of nature.'' We believe that, notwithstanding the
development of large-scale computer intensive models, there is still and
there will always be the need for simpler approaches. Any knowledge should
be seen as the collection of a hierarchy of descriptions, from the more
general level with few variables and fluxes, which like a cartoon
provides an outline of the main traits of the portrait, to the more
detailed microscopic approaches. In between, a series of intermediate
levels form the bridges between the two extreme modeling levels. Examples
are found in all sciences. For instance, in hydrodynamics, going from the
micro- to the macro-scales, we have the micro-level of molecular dynamics,
the density functional approaches, the Master equations, the Fokker-Planck
equations and finally the Navier-Stokes equation at the largest
coarse-grained scale.  The coarse-graining approach to modeling is based either
on averaging the micro-dynamics or (and this is the route followed here)
by using a ``Landau'' approach based on symmetry and conservation laws
that allows us to identify the leading variables and their interactions. 
As is well-known in statistical physics, this approach may introduce new terms that 
cannot be a priori predicted from the microscopic biological level, 
and which account in an effective emergent way for the many complicated pathways of interactions.   
It would be a fundamental mistake to expect that all terms can be explained or justified at the microscopic level. At a higher level, new terms and novel effective interactions should be
expected (Anderson, 1972).

In the sequel, we start from a simple but already
rather rich framework, that will provide a guideline for subsequent
developments.

\section{Mathematical formulation of the homeostasis
dynamical processes associated with the immune system}

\subsection{General formulation}

In order to formulate mathematically the various interactions between
cells and pathogens, we resort to the method of rate equations. The
method of rate equations is commonly employed for characterizing the
evolution of competing species. The general description of this approach
can be found in the book by Hofbauer and Sigmund (2002).
In the rate-equation approach, one considers the average
properties of a system and its constituents, with the averaging assumed
to be done over the whole body. Therefore, the spatial structure of the
latter can be arbitrary.

Our model can be viewed as a significant extension of simple
two-dimensional models of viral infection involving just normal cells
with population $N_1$ and altered cells with population number $N_2$.
One of the simplest representative of this family reads (Nowak, 2006)
\ba
\label{gw}
{d N_1 \over dt} &=& k - u N_1 - b N_1 N_2 \; ,    \\
\label{jhyl}
{d N_2 \over dt}  &=& b  N_1 N_2  - (u + v) N_2~,
\ea
where $N_1$ (resp. $N_2$) is the number of normal (resp. infected) cells.
The parameter $k$ represents the birth flux of normal cell, $u$ is
their normal death rate and the last term $ - b N_1 N_2$ in the first
equation is the rate of infection of normal cells when they come in
contact with altered cells. The destruction term for the population
$N_1$ is a creation term for the population $N_2$. The last term
$-(u + v) N_2$ in the second equation with  $v>0$ embodies
the increased mortality rate due to the alteration.
This model is sufficient to represent simple infection diseases,
with a competant immune system in the presence of circulating germs.

However, there is much more to the immune system dynamics than just
its response to external pathogens. The immune system is a complex
network of interacting components which also evolve and change,
accumulating history-dependent characteristics, properties, strengths
and deficiencies. Specifically, a useful model of the immune system
should be able to include in a single framework the four main classes
of clinical affections: allergies, chronic infections, auto-immune
diseases and cancers. The model we discuss below aims at presenting
a coherent system view of these different affections.

We consider five different agents constituting a living organism.
These are:
\begin{description}
\item[(i)] normal healthy cells, whose number will be denoted by
$N_1$;
\item[(ii)] altered, or infected, cells, whose number is $N_2$;
\item[(iii)] adaptive, or specific, immune cells, quantified by the
number $N_3$;
\item[(iv)] innate, or nonspecific, immune cells, with the number
$N_4$;
\item[(v)]  pathogens, their number being $N_5$.
\end{description}
The non-specific immune response includes polynucleus cells,
pro-inflammatory enzymes and the complement system. The specific
immune response includes the lymphocytes and their product antibodies.
Dividing the immune response into two components comes at the cost
of augmenting the complexity of the system, but seems necessary
to capture how inflammation and other non-specific
responses might lead to ill-adapted or even negative effects
leading to feedbacks on the specific part of the immune system,
possibly at the origin of allergies, chronic infections and auto-immune
diseases. However, we use below a parametrization which allows one to combine
the two components of the immune system into a single effective one, thus
decreasing the complexity of the system to four coupled ODEs. Differentiating
the specific interactions and differences between the two immune systems
will be the subject of another study.

The dynamics of the agents described above, and their interactions,
can be described similarly to those of other competing
species in the models of population evolution, using a system of ODEs obeying
the Markov property. The general rate
equations for the competing species thus read
$$
\frac{dN_i}{dt} = R_i N_i + F_i \; ,
$$
where $i=1,...,5$,
$R_i$ are rate functions, and $F_i$ are external influxes.
Rather than a constant incoming rate of normal cells as
in (\ref{gw}), we use a proportional growth term $R_1 N_1$, to
emphasize the endogenous nature of the dynamics. Higher order
nonlinear terms ensure saturation to a constant number of normal
cells. In contrast, using a constant incoming rate of healthy cells
relies on the existence of a reservoir, whose dynamics should not be
kept exogenous since our goal is to provide a self-consistent
classification of the different states of the immune system-pathogen
complex.

The rate functions are dependent of the agent numbers, so that
$R_i=R_i(N_1,N_2,N_3,N_4,N_5)$. Also, both $R_i$ and $F_i$ can depend on
time $t$. The rate functions, in general, can be modeled in different
forms, having any kind of nonlinearity prescribed by the underlying
process. The most often employed form for the
rate functions, which we shall also use, is
\be
R_i = A_i +\sum_j A_{ij} N_j \; ,
\label{gjpgtvw}
\ee
which can be seen as a Taylor expansion in powers of the agent numbers.
Since $A_1>0$ for the normal cells, the corresponding decay term
$-A_{11} N_1^2$ is nonlinear. Both terms taken together make endogenous
the homeostatic equilibrium of the normal cells, thus avoiding the need
for an external source. As discussed below, except in the case of
spontaneously proliferating infected cells, all other decay terms are
otherwise linear.

Generally, we could include as well indirect interactions with
higher-order nonlinearities in expression (\ref{gjpgtvw}). For instance,
the third-order terms $A_{ijk}N_iN_jN_k$ could be included. We do not
consider such terms for two reasons. First, such indirect interactions
are usually less important. Second, their influence, to a large extent,
has already been taken into account by a combination of several second-order
terms, such as $A_{ij}N_iN_j$. In the majority of cases, the above form of
the rate equations is quite sufficient for catching the main features of the
considered dynamics.

We are going to describe the leading processes and their associated
equations for the main terms of the rate equations. These terms
take into account all basic interactions between the constituents of a living
organism. The discussed direct interactions yield the nonlinear
terms of second order.

The structure of the kinetic equations must satisfy the following general rules:

\vskip 2mm

(i) {\it Same-order nonlinearities}. Nonlinear terms, characterizing
interactions between cells, have to be of the same order of nonlinearity.

\vskip 2mm

(ii) {\it Self-consistency of description}. Equations include the
major processes between cells. All interaction parameters are to
be treated on the same footing. The modulation of these parameters by
secondary processes must be either taken into account everywhere or
neglected everywhere.

\vskip 2mm

(iii) {\it Action-counteraction dichotomy}. Each process $A_{ij}N_iN_j$,
representing the interaction between an $i$-cell with a $j$-cell, must
have its counterpart process $A_{ji}N_jN_i$ (with $A_{ji}$ not
necessarily opposite to $A_{ij}$).

\subsection{Specific determination of the kinetic rates}

\subsubsection{Kinetics of healthy cells $N_1$}

{\parindent=0pt
(1) Healthy cells die with a natural decay rate and they are also
produced by a specialized system of cells. The net (total) linear rate
of death-birth of healthy cells is $A_1$ and the corresponding term is
$A_1N_1$.

(2) The decay of healthy cells is described by the nonlinear term
$-A_{11}N_1^2$.

(3) The adaptive immune system occasionally attacks healthy cells, which
provokes auto-immune diseases. This is described by the term
$-A_{13}N_1N_3$.

(4) The innate immune system sometimes also attacks healthy cells (e.g.
through inflammation), which
is represented by the term $-A_{14}N_1N_4$.

(5) Healthy cells are infected by pathogens, a process given by the
term $-A_{15}N_1N_5$.
}

\subsubsection{Kinetics of infected (or anomalous) cells $N_2$}

{\parindent=0pt
(1) Infected cells die with a natural rate $A_2$, hence the decay term
$-A_2N_2$ with $A_2>0$. For $A_2>0$, i.e., such that the term $-A_2N_2$
corresponds to a net death quotient, infected cells cannot significantly
duplicate themselves.

(1bis) It is also interesting
to consider, the case where infected cells
multiply with a net positive growth rate $|A_2|$ corresponding to choosing
$A_2<0$. As the analysis will show, this will allow us to
obtain states that can be interpreted as cancer afflictions.  For instance,
chronic inflammation (as in the case of
stomach lining) can lead to cancer (Helicobacter pylori infection can
lead to stomach ulcers short term and stomach cancer long term).
We also use the terminology ``cancer'' loosely to refer to situations
in which the multiplication of infected cells during chronic infections
occurs as if it was a cancer according to a dynamic process similar
to the cellular tumor growth.

(2) The decay process can include a nonlinear contribution with the term
$-A_{22}N_2^2$ with $A_{22} \geq 0$.

(3) Infected cells are killed by the adaptive immune system, thus
the term $-A_{23}N_2N_3$.

(4) They are also killed by the innate immune system, so the term
$-A_{24}N_2N_4$.

(5) The natural decay rate of infected cells can be increased by interacting
with pathogens, which implies the term $-A_{25}N_2N_5$.

(6) The number of ill cells increases by the infection of
healthy cells by pathogens, which is represented by the term
$A_{51}N_5N_1$.
}

\vskip 2mm

\subsubsection{Kinetics of the immune system}

The immune system of vertebrates can be divided into the
innate component (macrophages, neutrophils, and many types of proteins
involved in inflammation responses) and the adaptive immune system
(antibodies and lymphocytes B  and T). These two components interact
with each other, cooperating and regulating each other. Hormonal fluxes
and the metabolism of fatty acids (precursors of prostaglandins with
opposite effects) also play an important role in
the regulation of pro- or anti-inflamatory signals. We thus
specify the dynamics of the two immune system components as follows.

Recent research show that that `danger signals',
such as those found during viral infections, can be recognized by T and B cells
of the adaptive immune system (Marsland et al., 2005a,b). It was generally
considered that such sensing of `danger signals' was limited to cells of the
innate immune system. The fact that
T cells have evolved to recognize 'danger signals' opens a wide spectrum of
possibilities including novel mechanisms for the maintenance of immune memory,
the development of autoimmunity and general T cell homeostasis. We
take into account this phenomenon in our description of the activation
of the immune cells.

\vskip 0.3cm
\noindent {\bf Kinetics of the adaptive immune system $N_3$}

{\parindent=0pt
(1) Immune cells in the adaptive immune system die by apoptosis, which
is described by the term $-A_3N_3$.

(2) For generality, the nonlinear decay, with the term $-A_{33}N_3^2$
is included, though it may be subdominant.

(3) The activity of immune cells is supported by healthy cells, whose
part, the marrow, reproduces the immune cells; these processes are
described by the term $A_{31}N_3N_1$.

(4) Adaptive immune cells are activated by infected cells, which gives
the term $A_{32}N_3N_2$.

(5) The adaptive immune system can be inhibited by the innate part of the
immune system, which corresponds to the term $-A_{34}N_3N_4$.
For instance, allergic inflammation of the lung inhibits pulmonary
antimicrobial host defense (Beisswenger et al., 2006). This process may
be part of the control of the immune
response after the removal of the infection or of the traumatism in order to
avoid an auto-immune excess. More generally,
overwhelming activation of innate immunity during sepsis produces profound,
relatively long-lived depression of adaptive immunity, evidenced by
sepsis-induced increased apoptosis of lymphocytes which decimates the
population (Hotchkiss and Karl, 2003). Another, specific instance is
the evolution of bacterial ``escape mechanisms,'' such as activation
of innate immunity resulting in a cellular response that inhibits
adaptive immunity.  For example, macrophages infected with Mycobacterium
tuberculosis secrete interleukin-6 which inhibits the response of
neighboring (uninfected) macrophages to interferon-gamma
(Nagabhushanam et al., 2003).

(6) Adaptive immune cells are activated by pathogens, implying the
term $A_{35}N_3N_5$. In reality, the adaptive immune system is activated
by antigen presenting cells, such as dendritic cells, macrophages,
and other elements of the innate immune system. Because we would
like to account for both the triggering as well as inhibitory effect of
the innate immune system on the adaptive one, we have chosen to separate
its actions, so that the activation is here proxied as if by a direct
action of the pathogens, while the inhibition is explicitly written as
the direct interaction $-A_{34}N_3N_4$ previously discussed.
}

\vskip 0.3cm
\noindent {\bf Kinetics of the innate immune system $N_4$}

{\parindent=0pt
(1) The natural decay is given by the term $-A_4N_4$.

(2) The decay can be increased by the nonlinear term $-A_{44}N_4^2$,
which, though, may be quite small.

(3) The proliferation of immune cells, supported by healthy cells, is
characterized by the term $A_{41}N_4N_1$.

(4) The activation by damaged cells gives the term $A_{42}N_4N_2$.

(5) The innate immune system can be activated by the adaptive part of
this system, yielding the term $A_{43}N_4N_3$.
At the biological level, this activation can proceed via feedback loops.
The adaptive immune system secretes specific antibody  (IgE in classical
allergies, IgG4 for food intolerance). The complex AG-IgE (N4) activates
mastocytes, which themselves liberate mediators of the non-specific
inflammatory reaction (more than 300 proteases can be involved). The
latter repress the specific immune system via the negative feedback
of the Jayle cycle, leading to a deficit in antibodies, except for
the IgE and/or IgG4, which induce a positive feedback towards a stronger
inflammation response, via the degranulation of mastocytes and the
attraction of the eosinophils.

(6) Pathogens activate the innate immune system, hence the term
$A_{45}N_4N_5$.
}

\subsubsection{Kinetics of pathogens (or allergens, chemical or
ionised particles) $N_5$}

{\parindent=0pt
(1) Pathogens have a finite lifetime, thus the existence of the term
$-A_5N_5$, where $A_5$ is the ``clearance rate'' of the pathogens.

(2) In general, the nonlinear  death (or elimination) term
$-A_{55}N_5^2$ can also exist.

(3) By the action-counteraction principle that we follow
for the coarse-grained description of the dynamics of the
five compartments, the existence of the term
$A_{25}N_2N_5$ in the equation determining the rate $dN_2/dt$
of change of the number of infected cells should be accompanied by
a term $A_{52}N_5N_2$ entering in the equation for $dN_5/dt$.
An attempt for a possible biological interpretation is as follows.
When infected cells die by lysis catalyzed by the presence of
pathogens, they release the pathogens they contained.
A mass action description is then given by the term
$A_{52}N_5N_2$. One could argue that the rate of production
of pathogens due to lysis of infected cells should be better
described by a term proportional to the number $N_2$
of infected cells, to the rate of lysis and to the number of
infectious virus particles produced by the infected cell upon lysis. But consistent
with our modeling strategy, we keep the above term $A_{52}N_5N_2$
as a coarse-grained description of how the interactions between
the infected cells and the pathogens affect the dynamics of the later.

(4) Pathogens are killed by the adaptive immune system, giving the term
$-A_{53}N_5N_3$.

(5) They are also killed or hindered by the innate immune system,
leading to the term $-A_{54}N_5N_4$.

(6) There exists a continuous supply of pathogens into the organism
from the exterior, represented by the influx $F$, which is, in general,
time varying (but here we will take this flux constant in our analysis of
stationary equilibrium states).
}

\subsection{Kinetic equations}

Summarizing the processes described above, we obtain the following system of
kinetic equations
\be
\label{kin1}
\frac{dN_1}{dt} = A_1N_1 - A_{11}N_1^2 - A_{13}N_1N_3 -
A_{14}N_1N_4 - A_{15}N_1N_5  \; ,
\ee
\be
\label{kin2}
\frac{dN_2}{dt} = - A_2N_2 - A_{22}N_2^2 - A_{23}N_2N_3 -
A_{24}N_2N_4 - A_{25}N_2N_5 + A_{51}N_5N_1 \; ,
\ee
\be
\label{kin3}
\frac{dN_3}{dt} = - A_3N_3 - A_{33}N_3^2 + A_{31}N_3N_1 +
A_{32}N_3N_2 - A_{34}N_3N_4 + A_{35}N_3N_5  \; ,
\ee
\be
\label{kin4}
\frac{dN_4}{dt} = - A_4N_4 - A_{44}N_4^2 + A_{41}N_4N_1 +
A_{42}N_4N_2 + A_{43}N_4N_3 + A_{45}N_4N_5  \; ,
\ee
\be
\label{kin5}
\frac{dN_5}{dt} = - A_5N_5 - A_{55}N_5^2 + A_{52}N_5N_2 -
A_{53}N_5N_3 - A_{54}N_5N_4  + F \; .
\ee

{\bf Remark}: The two equations for the adaptive and innate
immune compartments are structurally the same, except for one important
feature, namely the difference in the sign of their mutual interactions.
\begin{itemize}
\item The term $- A_{34}N_3N_4$ in the dynamics of $\frac{dN_3}{dt}$ expresses
(for $A_{34}>0$) a repression of the active immune cells induced by the
innate system. This feedback of $N_4$ on $N_3$ occurs with a delay
whose deficiency may cause auto-immune diseases.

\item In contrast, the term $+ A_{43}N_4N_3$ in the dynamics of
$\frac{dN_4}{dt}$ expresses (for $A_{43}>0$) the activation of the innate
system by the active immune cells.  This corresponds to a vulnerability
towards allergies. The reverse sign $A_{43}<0$ also occurs (not considered
here) due to the high
specificity of the adaptive cells $N_3$ which, by reducing the infections,
may inhibit the activation of the innate immune system.
\end{itemize}

However, since the actions of both
adaptive and innate components on the other cell types $N_1, N_2$ and $N_5$
are structurally and qualitatively identical, the presence or
absence of these different regulations are not
expected to lead to significant differences in the obtained classification
of illnesses. One can expect that, at our present level of description
of the dynamics of the $5$ cell types, the two immune system components
can be combined into a single one as they act qualitatively as a single
effective system.

\subsection{Reduction to dimensionless quantities}

Since the cell numbers can be extremely large, it is convenient to
work with normalized quantities, reduced to a normalization constant
$N$ (which is {\it not} the total number of cells of the organism). For this purpose, we introduce the 
normalized cell numbers
\be
\label{cellfr}
x_i \equiv \frac{N_i}{N} \qquad (i=1,2,3,4,5)\; .
\ee
It is also convenient to deal with dimensionless quantities for decay
rates and interaction parameters and to measure time in relative units.
Let the latter be denoted by $\tau$. Then the dimensionless decay rates
are given by
\be
\label{decrate}
\al_i \equiv A_i \tau
\ee
and the dimensionless interaction parameters by
\be
\label{inparam}
a_{ij} \equiv A_{ij} N \tau \; .
\ee
The dimensionless pathogen influx is given by
\be
\label{pathinf}
\vp \equiv \frac{\tau}{N}\; F \; .
\ee
Using these notations, and measuring time in units of $\tau$, we come
to the system of dimensionless kinetic equations in the form
\be
\label{dimeq}
\frac{dx_i}{dt} = f_i \; ,
\ee
where $x_i$ are the normalized cell numbers (\ref{cellfr}) and the right-hand sides are
\be
\label{f1}
f_1 = \al_1 x_1 - a_{11}x_1^2 - a_{13}x_1 x_3 - a_{14}x_1 x_4 -
a_{15}x_1 x_5  \; ,
\ee
\be
\label{f2}
f_2 = - \al_2 x_2 - a_{22}x_2^2 - a_{23}x_2 x_3 - a_{24}x_2 x_4 -
a_{25}x_2 x_5 + a_{51} x_5 x_1\; ,
\ee
\be
\label{f3}
f_3 = - \al_3 x_3 - a_{33}x_3^2 + a_{31}x_3 x_1 + a_{32}x_3 x_2 -
a_{34}x_3 x_4 + a_{35}x_3 x_5  \; ,
\ee
\be
\label{f4}
f_4 = - \al_4 x_4 - a_{44}x_4^2 + a_{41}x_4 x_1 + a_{42}x_4 x_2 +
a_{43}x_4 x_3 + a_{45}x_4 x_5  \; ,
\ee
\be
\label{f5}
f_5 = - \al_5 x_5 - a_{55}x_5^2 + a_{52}x_5 x_2 - a_{53}x_5 x_3 -
a_{54}x_5 x_4 + \vp \; .
\ee
These are the main equations for the organism homeostasis we shall
analyse. The behavior of the dynamical systems of such a high
dimensionality can possess quite nontrivial features (Arneodo et al.,
1980; Hofbauer and Sigmund, 2002; Ginoux et al., 2005).

\subsection{A priori consideration}

This system is reminiscent of two preys-two predators systems (see e.g.
(Hsu et al., 2001; Xiang and Song, 2006)). The healthy cells $N_1$
corresponds to ``food'' or to a first ``prey''  ``hunted'' by the immune
cells (in case of auto-immune disorder tendency $a_{13} >0$, $a_{14}>0$)
and a fifth species, the pathogens. The infected cells $N_2$ corresponds to
a second prey hunted also by the immune cells. The two types of immune
cells $N_3$ and $N_4$ are the predators, which in addition interact
directly through repressive-promoting asymmetric interactions. Note that the predation
or cytotrophic mechanism of the immune system is a priori beneficial for
the organism, ensuring the renovation of tissues by elimination of
aging or damaged cells. The crucial
novel feature is the presence of the fifth species, the pathogens, which play
a rather complex mixed role: it is like a predator of the first prey
$N_1$, while it is a catalyst of food intake for the second prey $N_2$ as
well as for the two predators. Previous studies of one prey-two predators
and of two preys-two predators have exhibited very rich phase diagrams.
Having a fifth pathogen component, we expect even more complex dynamics,
and our finding of a rather strong structural stability presented below
comes as a surprise.

The dynamical model, defined by Eqs. (\ref{dimeq})-(\ref{f5}), is a
five-dimensional dynamical system with 29 parameters. It seems, at
first glance, that such a large number of parameters, which are not
strictly defined, makes it impossible to extract reasonable conclusions
from such a complex model. However, fortunately, the qualitative
behavior of dynamical systems often depends mostly, not on the absolute
values of the control parameters, but rather on their signs. As
emphasized by Brown et al. (2003, 2004), dynamical systems modeling
complex biological systems always have a large numbers of poorly known, or
even unknown, parameters. Despite of this, such systems can be used
to make useful qualitative predictions even with
parameter indeterminacy. This reflects the fact that models which have been
constructed on the basis of good physical and biological guidelines
enjoy the property that their qualitative behavior is weakly influenced
by the change of the parameter values within finite bounds  (Brown et al.,
2003, 2004).

In the language of the theory of dynamical systems (Scott, 2005), this
property is known as {\it structural stability}. The analysis of our model
(\ref{dimeq})-(\ref{f5}) performed by varying the control parameters over
wide ranges confirm the existence of structural stability. It turned out
that the dynamical system (\ref{dimeq})-(\ref{f5}) is remarkably structurally
stable, always displaying four stable stationary states
for all parameter values.  However, the nature and properties of some
of these four states change,  consistent with the existence of different illnesses.

We now describe in detail the results obtained respectively for $A_2>0$
(infected cells tend to die) and $A_2<0$ (infected cells tend
to proliferate). In each of these two
cases, we also consider what can be considered in some sense two opposite
regimes, respectively without and with an auto-immune attack of immune
cells on normal cells. Interesting coexistence and boundaries between
different states characterize possible paths from health to illness
(allergies, chronic infections, auto-immune diseases, cancer), critical
illness, and death.

\section{Existence and stability of stationary states with decaying ill
cells ($A_2>0)$
\label{jkhkybgr}}

\subsection{Setting of the parameters}

In general, the analysis of the stationary states and their stability
for a five-dimensional dynamical system, such as given by Eqs.
(\ref{dimeq})-(\ref{f5}), can only be accomplished numerically. In
order to simplify the representation, we assume the following features
that are justified by the medical literature at the basis of our model.
First, we take the same apoptosis rates of both compartments of the
immune system, setting
\be
\label{al}
\al \equiv \al_3 = \al_4 \; .
\ee
This parameter $\alpha$ can be arbitrary. Other cell rates will be taken with
\be
\label{al125}
\al_1 = \al_2 = \al_5 = 1 \; .
\ee
The nonlinear decay plays an important role only for healthy cells,
since the term $a_{11}$ limits the growth of the body, which, if
$a_{11}$ was zero, would grow without bounds due to the positive
$\al_1$. Because of this, we fix
\be
\label{a11}
a_{11} = 1 \; .
\ee
At the same time, the nonlinear decay of other cells can be neglected,
since their linear decay rates are already negative and thus should
dominate. Indeed, when a linear term is negative, it leads to a natural
exponential decay. The negative nonlinear term is then not necessary.
Technically, its presence would just modify a little the quantitative values
of the normalized cell numbers, without impacting on the qualitative structure of the
phase diagrams. So, we take
\be
\label{aii}
a_{22} = a_{33} = a_{44} = a_{55} = 0 \; .
\ee
To be able to derive analytically at least some of the formulas, we
set, for simplicity,
$$
a_{15} = a_{23} = a_{24} = a_{25} = a_{51} = a_{32} = a_{34} =
$$
\be
\label{aij}
= a_{35} = a_{42} = a_{43} = a_{45} = a_{52} = a_{53} =
a_{54} = 1\; .
\ee
The two opposite cases presented below differ from each other by
whether or not the immune system attacks healthy cells, causing
autoimmune diseases.

\subsection{No auto-immune disorder \label{jhkke}}

If the immune system attacks only the infected cells and the
pathogens, but does not attack healthy cells, then
\be
\label{a13}
a_{13} = a_{31} = a_{14} = a_{41} = 0 \; .
\ee
In this case, the right-hand sides of the dynamical system (\ref{dimeq})
become
$$
f_1 = x_1 ( 1 - x_1 - x_5)\; , \qquad
f_2 = -x_2 ( 1 + x_3 + x_4 + x_5) + x_1x_5 \; ,
$$
$$
f_3 = x_3 ( x_2 - x_4 + x_5 -\al)  \; , \qquad
f_4 = x_4 ( x_2 + x_3 + x_5 -\al)  \; ,
$$
\be
\label{dineq1}
f_5 = x_5 ( x_2 - x_3 - x_4 -1) + \vp  \; .
\ee

For what follows, it is convenient to introduce the total normalized cell numbers of
immune cells
\be
\label{y}
y \equiv x_3 + x_4 \; .
\ee
In a remark after equations (\ref{kin1}-\ref{kin5}),
we already alluded to the fact that  the two immune system components
could be combined into a single one as they act qualitatively as a single
effective system on the other species $N_1, N_2$ and $N_5$. Here, this
holds quantitatively as there is an exact cancellation of the presence of
the $-x_3 x_4$ term in $f_3$ by the corresponding $+x_3 x_4$ term in
$f_4$, which together with the symmetry between $x_3$ and $x_4$ in the
other equations, ensures that the dependence on $x_3$
and $x_4$ in all equations for $dx_1/dt, dx_2/dt, dy/dt$ and $dx_5/dt$ only
appear via their sum variable $y$ defined in (\ref{y}).
Of course, this exact cancellation of the $x_3 x_4$ term in $f_3$ and $f_4$
and the symmetric role of $x_3$ and $x_4$ only
occur due to the special symmetric choice of the system parameters in Eq.
(\ref{aij}). Only with such a symmetry, can the immune system be treated
as a total part of the organism,
without separating it into the innate and adaptive components. The
difference between the latter arises if the parameters, characterizing the
parts of the immune system, differ from each other. In real life,
these  parameters are probably slightly different, thus breaking the symmetry
between the components of the immune system. Many clinical disorders,
and in particular many allergic manifestations, are underpinned by 
some dynamical asymmetry, i.e., by a delay in the feedback exerted by one 
immune component onto the other. This may explain why episodic acute
events are not chronic disorders as, after a while, the symmetry is restored. But, for the sake of
simplicity, we preserve for a while this symmetry, which allows us to
reduce the five-dimensional dynamical system to a four-dimensional
one.

The symmetric choice of the system parameters in Eq. (\ref{aij}) allows us,
for the time being, to make no distinctions between the parts of the
immune system. Then, considering the stationary states which are
classified below, we have the
equivalence of two cases, when $y^*=x_3^*$, while $x_4^*=0$, or when
$y^*=x_4^*$, but $x_3^*=0$. This follows directly from the consideration
of the corresponding five-dimensional dynamical system with the
``forces'' given by (\ref{dineq1}).

The stationary states are given by the solutions of the equations
$f_i=0$, in which we assume the pathogen flux $\vp$ to be constant.
The stability of these states is investigated by means of
the Lyapunov stability analysis (see Appendix A). For this purpose, we
calculate the Jacobian matrix $[J_{ij}]$, with the elements $J_{ij}=
\partial f_i/\partial x_j$, find its eigenvalues, and evaluate the
latter at the corresponding fixed points. All that machinery is rather
cumbersome, and we present only the results, omitting intermediate
calculations. We find four stable stationary states ($A, B, C$ and $D$).

Figure 1 presents the phase diagram, or domain of stability of each
of the four states, which we denote as $A$, $B$, $C$, and $D$, in the
parameter plane ($\alpha,\vp$). These four states are
denominated without indices to distinguish them from the other cases
studied below in which we consider the possibility for auto-immune
(condition (\ref{a13}) is not met) or cancer ($A_2<0$)
tendencies. This diagram is obtained as follows.

\vskip 3mm

{\bf State $A$} (strong active immune system)

\vskip 2mm

The domain of stability of  State $A$ is defined by the union of the two
sets of conditions:
\be
\label{alfi1}
0 < \al < \frac{1}{2} \; , \qquad \vp > \al (1 - \al) \; ,
\ee
\be
\label{alfi2}
\frac{1}{2} < \al <  1 \; , \qquad \vp > \al^2 \; .
\ee
In other words, State A occurs when either conditions (\ref{alfi1}) or
(\ref{alfi2}) hold. In a nutshell, these conditions require that (1) the
immune system is characterized by a relatively small apoptosis rate
($\alpha <1$) and (2) the flux $\vp$ of pathogens should not be too small.

The solutions expressing the values of the numbers of cells corresponding
to the fixed point state $A$ are extremely cumbersome, and we do not present
their analytical expressions but rather their graphical dependence in
Figure 2 as functions of the apoptosis rate $\alpha$ for three different
values of the pathogen flux $\vp$. Increasing the apoptosis rate reduces the
normalized cell numbers of healthy cells because the reduced number of immune
cells imply a less effective defense against the pathogens.
State $A$ can survive under arbitrary high pathogen
influx, as the immune system is very strong.

For a large pathogen influx, $\vp>1$, the stationary cell numbers are well
represented by their asymptotic forms
$$
x_1^* \; \simeq \; (1-\al) + (1-\al) \;\frac{\al^2}{\vp} \; - \;
2 (1-\al)^2 \; \frac{\al^3}{\vp^2} \; ,
$$
$$
x_2^* \; \simeq \; (1-\al)\; \frac{\al^2}{\vp} \; - \;
2(1-\al)^2 \; \frac{\al^3}{\vp^2} \; ,
$$
$$
y^* \; \simeq \; \frac{\vp}{\al} \; - \; \al - (1-\al)\left ( 1-2\al
\right )\; \frac{\al}{\vp} + (1-\al)^2 (2-5\al)\; \frac{\al^2}{\vp^2} \; ,
$$
\be
\label{ass1}
x_5^* \; \simeq \; \al \; - \; (1-\al)\; \frac{\al^2}{\vp} +
2(1-\al)^2 \; \frac{\al^3}{\vp^2} \; .
\ee
We thus have, for $\vp \to +\infty$,
\be
x_1^* \to 1-\alpha~, ~~~x_2^* \to 0~,~~~
y^* \to {\vp \over \alpha}~,~~~x_5^* \to \alpha~.
\label{wrtbrt}
\ee
These asymptotic expressions embody the remarkable observations
that can be made upon examination of Figure 2, namely the very weak
dependence of $x_1^*$ and $x_5^*$ as a function of $\vp$, coming
together with the very large dependence of $y^*$ and to a lesser extent
of $x_2^*$ as a function of $\vp$. Indeed, we see in expression (\ref{wrtbrt})
that, as the organism is subjected to an increasing pathogen concentration
$\vp$, the immune response blows up proportionally ($y^* \to \vp/\alpha$),
ensuring an almost independent concentration of healthy cells. This strong
response of the immune system has the concomitant effect of putting at bay the
pathogens whose concentration $x_5^*$ is very weakly depending on $\vp$.
This strong immune system response has also the rather counter-intuitive
consequence that the number $x_2^*$ of infected cells is a decreasing function
of the pathogen flux $\vp$.

In summary, State A describes an organism with a ``healthy'' immune system,
capable of controlling any amount of exogenous flux of pathogens. This occurs
only for a sufficiently small apoptosis rate ($\alpha <1$) and under the
influence of a sufficiently large pathogen flux $\vp$. The later condition
is a vivid embodiment of our conjecture formulated in Section 2 that a
healthy homeostasis state requires a sufficiently strong pathogen flux;
in other words, health is not the absence of pathogens, but rather a strong
ability to find balance by counteracting any pathogen attack.

When the flux $\vp$ of pathogens decreases below the limits given by
conditions (\ref{alfi1}) or (\ref{alfi2}), an interesting phenomenon appears:
State A with a very active immune system is replaced by State B with an
evanescent immune system. In other words, there is a critical threshold for
the pathogen flux $\vp$ below which the immune system collapses.

\vskip 3mm

{\bf State $B$} (evanescent immune system)

\vskip 2mm

The second State $B$ is characterized by the following stationary
cell numbers:
$$
x_1^* = 1 + \frac{1-\vp}{3Z} \; - \; Z \; , \qquad
x_2^* = 1 \; -\;  \frac{3\vp Z}{3Z^2-(1-\vp)} \; ,
$$
\be
\label{fract}
x_3^*= 0 \; , \qquad  x_4^* = 0 \; , \qquad y^* = 0 \; , \qquad
x_5^* = Z \; - \; \frac{1-\vp}{3Z} \; ,
\ee
in which $Z=Z(\vp)$ is defined as
\be
Z \equiv \left [ \frac{\vp}{2} + \sqrt{\frac{1}{27}(1-\vp)^3 +
 \frac{\vp^2}{4}} \right ]^{1/3} \; .
 \label{hjbepg}
\ee
The expressions of
$x_1^*, x_2^*$ and $x_5^*$ in (\ref{fract}) can approximately be written as
\be
\label{approxB2}
x_1^* \cong 1 -\vp \; , \qquad x_2^* \cong (1-\vp)\vp \; ,
\qquad x_5^* \cong \vp \; .
\ee
Note that none of the cell numbers depend on the apoptosis rate $\alpha$,
due to the absence of immune cells.
State $B$ is an organism with vanishing immune cells, but which manages
to survive solely from its regenerative
power when the pathogens are not too virulent.
The behavior of the non-zero
stationary cell normalized cell numbers for State $B$ as functions of the pathogen flux $\vp$
is shown in Fig. 3.

State $B$ has the following domain of existence:
\be
\label{stabB}
\al > x_2^* + x_5^* \; , \qquad 0 \leq \vp \leq 1 \; ,
\ee
which can be well approximated by
\be
\label{approxB1}
\al > (2-\vp)\vp \; , \qquad 0\leq \vp \leq 1 \; .
\ee

Consider an organism which starts in State A with a strong responsive
immune system with some fixed $\alpha <1$. Suppose that, for some reason,
the pathogen flux decreases so that the boundary $(2-\vp)\vp = \alpha$
is crossed and the organism evolves to State B (see Figure 1). Comparing
the two states, we see that State B here corresponds to an organism that
has no responsive immune cells, because the immune system is
insufficiently stimulated by pathogens. This may seem okay as long as
$\alpha$ remains smaller than $1$, as a stronger pathogen flux will
suddenly trigger the immune response and shift the organism back to State
A. However, something bad can happen from this lack of stimulus: for
larger apoptosis rates $\alpha >1$, the organism manages to survive in
State B as long as the pathogen flux $\vp$ remains small. But would a
burst of pathogen influx occur, State B would be replaced abruptly by
death (State D discussed below), without the immune system being able
to do anything. Thus, State B cannot be considered healthy as it is vulnerable
to a change of pathogen flux that may have catastrophic consequences.

Structurally,  the stability of State $B$ relies on a condition which is the
qualitative antinomy of the condition of stability for State $A$. Indeed,
State $B$ exists only for sufficiently small fluxes of pathogens and large
apoptosis rate (weak immune system). In contrast, State $A$ can absorb any
pathogen flux, given that the apoptosis rate is sufficiently small
(sufficiently strong immune system).  State $B$
describes for instance  the situation of cancer patients after
chemotherapy: the chemotherapy accelerates cell death, a weakened
immune systems results, and the patients frequently develop sepsis
spontaneously, usually with their own organisms/bacteria.  Death
rates are relatively high.

\vskip 3mm

{\bf State $C$} (critically ill, $1 < \alpha < \vp $)

\vskip 2mm
In State $C$, the stationary cell numbers are
\be
\label{fractC}
x_1^* = 0 \; , \qquad x_2^* = 0 \; , \qquad
y^* = \frac{\vp}{\al} \; - \; 1 \; , \qquad x_5^* = \al \; .
\ee
There are no healthy cells, though the immune system still fights pathogens.
The domain of stability of State $C$ is given by the inequality
\be
\label{stabC}
1 < \al < \vp \; .
\ee
State C is reached from State A, when the pathogen flux is large,
by weakening the immune system (i.e., increasing
the apoptosis rate $\alpha$). When the boundary $\alpha=1$ is
reached, State A is replaced by State C: the number of normal cells
as well as the number of infected cells collapse. It is as if the
organism would put all its energy on the proliferation of the immune cells
to fight the pathogens. The immune system is able to stabilize the number
of pathogens to a fixed value $\alpha$ for any pathogen flux $\vp > \alpha$,
solely controlled by the apoptosis rate $\alpha$.

Figure 4 shows
indeed that the number of immune cells increases with the pathogen flux $\vp$
and can be very large if $\alpha$ is not too large. State C can be interpreted
as a critically-ill state, where the organism's survival is dependent
upon the successful eradication of the pathogens, diverting if necessary
energy from other activities.

The organism in the critically ill State $C$ can recover by strengthening
its immune system, i.e., by decreasing the apoptosis rate $\alpha$ below
the boundary value $1$, leading to a transition back to State $A$ characterized
by a non-zero number of normal cells, hence a partial recovery, in the
presence of the strong flux of pathogens.

Interestingly, in State $C$, it would be lethal to decrease the flux
$\vp$ of pathogens at fixed apoptotis rate $\alpha$, as this would lead
the organism to the death State $D$ discussed below.
Perhaps, this suggests that it is more appropriate to stimulate the
immune system of critically-ill patients than to try to ensure a
pathogen-free environment. It is the still strong pathogen flux which
somehow maintains the organism alive, by stimulating
its weak immune system. Decreasing the pathogen flux removes any
stimulation and lead to the death State $D$, that we now describe.

We conjecture that State $C$ may provide an approximate description
of the so-called ``critically ill'' state of patients in intensive care
units, suggesting the interpretation that the dynamics of the immune
system in critical illness is a form of ``self-destruct.'' This raises the
question of whether clinicians can cheat death by trying to avoid
programmed self-destruction?

\vskip 3mm

{\bf State $D$} (death, $1 < \vp < \al$)

\vskip 2mm

State $D$ is characterized by a vanishing number of living body cells and
only pathogens are present:
\be
\label{fractD}
x_1^* = 0 \; , \qquad x_2^* = 0 \; , \qquad
x_3^* = x_4^* = y^* = 0 \; ,\qquad x_5^* = \vp \; .
\ee
State $D$ is stable in the domain
\be
\label{stabD}
1 < \vp < \al \; .
\ee

\subsection{Auto-immune disorder}

Rather than imposing condition (\ref{a13}), we now consider the regime where
\be
\label{a13II}
a_{13} = a_{31} = a_{14} = a_{41} = 1 \; .
\ee
All the other parameters are set as in the previous section \ref{jhkke}.
The right-hand sides of the kinetic equations (\ref{dimeq}) now read
$$
f_1 = x_1 ( 1 - x_1 - x_3 - x_4 - x_5)\; , \qquad
f_2 = -x_2 ( 1 + x_3 + x_4 + x_5) + x_1x_5 \; ,
$$
$$
f_3 = x_3 ( x_1 + x_2 - x_4 + x_5 -\al)  \; , \qquad
f_4 = x_4 ( x_1 + x_2 + x_3 + x_5 -\al)  \; ,
$$
\be
\label{eqII}
f_5 = x_5 ( x_2 - x_3 - x_4 -1) + \varphi  \; .
\ee
The symmetric choice of the system parameters in Eqs. (\ref{aij}) and
(\ref{a13II}) together with the exact cancellation of the $x_3 x_4$ term
in $f_3$ and $f_4$ ensures that we can again use the total normalized number
$y$ of immune cells defined in (\ref{y}).

The analysis shows  that there are again only four stable stationary states.
Three of them, $B, C$ and $D$, are identical to those found in the previous
section \ref{jhkke}, with however distorted domains of stability for $B$
and $C$. The death state $D$ is identical in its characteristics and
domain of stability. The main change is that State $A$, characterized by
a strong active immune system, is changed into a new state that we call
$A_{\rm aut}$. Figure 5 presents the phase diagram, or domain of stability
of each of the four states ($A_{\rm aut}$, $B$, $C$, and $D$) in the
parameter plane ($\alpha,\vp$). One can see that the new state $A_{\rm aut}$
has a reduced domain of stability compared with State $A$, reflecting the
effect of the auto-immune attack of normal cells by immune cells.

\vskip 3mm

{\bf State $A_{\rm aut}$} (strong immune system with auto-immune disorder)

\vskip 2mm

This State is stable in the domain defined by
\be
\label{stabAII}
0 < \al < 2(2-\sqrt{2})=1.172\; , \qquad \vp_1 < \vp < \vp_2 \; ,
\ee
where $ \vp_1$ and $\vp_2 $ are given by
$$
\vp_1\equiv \max \left\{ 0,\; \frac{1}{2}(2-\al)\left ( 2-\al -
\sqrt{\al^2-8\al+8}\right )\right \} \; ,
$$
\be
\label{lim}
\vp_2\equiv \min \left\{ (2-\al)\al, \; \frac{1}{2}(2-\al)
\left ( 2-\al +\sqrt{\al^2-8\al+8}\right )\right \} \; .
\ee
Note that, in contrast with State $A$, State $A_{\rm aut}$ is
not able to sustain a virulent attack of pathogens since it evolves
into the critically ill state $C$ under large $\vp$ (see below). This occurs
notwithstanding the conditions of moderate to small apostosis rates $\alpha$
which are similar to those governing State $A$.

The difference between State $A_{\rm aut}$ and State $A$ is
exemplified in the dependence of the cell numbers associated with the fixed
point $A_{\rm aut}$ as a function of $\alpha$ and $\vp$:
$$
x_1^*= \frac{1}{2}\left [ \al + 2
-\sqrt{(2-\al)^2+\frac{8\vp}{2-\al}} \right ]\; ,
$$
$$
x_2^*= \frac{1}{2}\left [ \al - 2 - \frac{2\vp}{2-\al}
+ \sqrt{(2-\al)^2+\frac{8\vp}{2-\al}} \right ]\; ,
$$
$$
y^*= \frac{1}{2}\left [ \sqrt{(2-\al)^2+\frac{8\vp}{2-\al}}
\; -\al \; - \; \frac{2\vp}{2-\al} \right ]\; ,
$$
\be
\label{fixpA}
x_5^* = \frac{\vp}{2-\al} \; .
\ee
These expressions are illustrated in Figure 6 which shows
the behavior of the stationary solutions (\ref{fixpA}) as functions of
the apoptosis rate $\al$ for different pathogen influx $\vp$.
The main difference with State $A$ shown in Figure 2 is found
in the behavior of the number $x_1^*$ of normal cells. In Figure
2 for State A,  $x_1^*$ is a decreasing function of $\alpha$, while in
Figure 6 for State $A_{\rm aut}$,
$x_1^*$  is an increasing function of $\alpha$. This point is particularly
well demonstrated by the asymptotic values of expressions (\ref{fixpA})
for a vanishing pathogen flux $\vp \to 0$:
\be
x_1^* \to \alpha~,~~~x_2^* \to 0~,~~~y^* \to 1-\alpha~,~~~x_5^* \to 0~.
\label{hjlombrt}
\ee
The expressions (\ref{hjlombrt}) exhibit two characteristics of an auto-immune
disorder, which cripples the organism even in quasi-absence of
an external flux of pathogens.
First, for a very strong immune system (small apoptosis rate $\alpha$),
the organism is mostly active through its immune cells, while normal or
infected cells have disappeared and free pathogens are absent.
Second, the number of normal cells recovers
only for a sufficiently weak immune system ($\alpha \to 1)$.
A third characteristic of an auto-immune disease is observed in the lower left
panel of Figure 6, which shows a very weak dependence of the number
$y^*$ of immune cells as a function of the pathogen flux $\vp$. This results
from the fact that the immune cells do not react anymore to just the pathogen
and infected cells but also to the normal cells. They thus develop a balance
controlled endogenously within the organism,
that is weakly influenced by the pathogens.

The origin of these characteristics of  State $A_{\rm aut}$ is
obviously found in the fact
that immune cells tend to also attack normal cells, and therefore a
larger apoptosis rate is favorable for the survival of these normal cells.
But this state can only hold up for not too large pathogen flux rates:
the immune cells, whose population needs to be controlled by
apoptosis if the normal cells are not to be decimated, would be overwhelmed otherwise.

In summary,  State $A_{\rm aut}$ has clear characteristics of
an auto-immune disease, aided or catalyzed by pathogens.

\vskip 3mm

{\bf State $B$} (evanescent immune system)

\vskip 2mm
The expressions of the cell numbers in the State $B$ found here are 
identical to those (\ref{fract}) and (\ref{hjbepg}) reported
in Section \ref{jhkke}. The graphical representation of the stationary
solutions for State $B$ is
the same as for solutions (\ref{fract}) in Fig. 3.

The only difference lies in the domain of stability which now read
\be
\label{stabBII}
\al > 1 + x_2^* \; , \qquad 0 \leq \vp \leq 1 \; .
\ee
The boundary between States $A_{\rm aut}$ and $B$ is given by the line, in the
$(\alpha,\vp)$ plane, of equation
\be
\label{lineAB}
\al = 1 + x_2^*(\vp) = 1 + \vp -\left ( \frac{\vp}{2-\al}\right )^2 \; .
\ee
For $\vp < 1$, the stability conditions can be approximately represented as
\be
\label{apprBII}
\al > 1 + (1-\vp)\vp \; , \qquad 0 \leq \vp \leq 1 \; .
\ee
As in Section \ref{jhkke}, State $B$ is an organism with no immune cells, but
which manages to survive in the presence of pathogens, solely from its
regenerative power when the pathogens are not too virulent.
Notwithstanding the propensity for immune cells to attack normal cells
considered in the present section, State $B$ has no auto-immune disease
due to the complete neutralization of immune cells. As a consequence, the organism
is entirely subjected to the whims of the pathogen fluxes which control
the number of normal cells and of infected cells in the presence of the
regenerative power of the normal cells, as seen from expressions
(\ref{fract}) and (\ref{hjbepg}).

A novel feature is the existence of a ``re-entrant'' transition in the
range $1 < \alpha <  \alpha_{\rm re} \approx 1.25$. For a fixed
apoptosis rate $\alpha$ in this range, a very small pathogen
flux $\vp$ puts the organism in  State $B$. Increasing $\vp$
leads to the first crossing of the boundary (\ref{lineAB}): the increase
of the pathogen flux stimulates the immune system which then
generates a non-zero number of immune cells. As a consequence, the
pathogens are better combatted at the cost of a slight auto-immune
affliction (State $A_{\rm aut}$). A further increases of $\vp$ pushes
the organism back to State B, and then to the death State D.

States $A_{\rm aut}$ and $B$ illustrate a trade-off between either
having an auto-immune disease aided by pathogens or being apparently
cured of the auto-immune disease, at the cost of the neutralization of
the immune cells which makes the organism vulnerable to exogenous
fluxes of pathogens. Indeed, increasing $\vp$ above $1$ leads to the
death State $D$ as discussed below.

\vskip 3mm

{\bf State $C$} (critically ill)

\vskip 2mm

The expressions of the cell numbers in the State $C$
found here are identical to those (\ref{fractC}) reported
in Section \ref{jhkke}, with no normal or infected cells. The
organism has only immune cells fighting the pathogens.
State $C$ is an armed balance between the immune cells
and pathogens, at the expense of the normal and infected cells.

The difference with Section \ref{jhkke} is that the domain of
stability of State $C$ is much wider, since the organism now suffers
attacks from both immune cells and pathogens. The domain
of stability of State $C$ is defined by the inequalities
\be
\label{stabCII}
0 < \al < 1 \; , \qquad \vp > \al (2-\al)
\ee
and by the conditions
\be
\label{condC}
1 < \al < \vp \; .
\ee
The transition from State $A_{\rm aut}$ to State $C$ as $\vp$ increases
and crosses the line $\vp = \al (2-\al)$
illustrates the run-away effect of an auto-immune disease
occurring with a strong immune system which attacks the normal cells.
In the absence of the auto-immune mechanism, the critical state $C$
is reached only for a sufficiently weak immune system and a sufficiently
large pathogen flux (case of Section  \ref{jhkke}). Here,
in the presence of the auto-immune effect, the critically ill state occurs
for arbitrary strong immune systems, as soon as the pathogen flux is
larger than the threshold $\vp = \al (2-\al)$. In the presence of the
auto-immune effect, the organism has no solution but to surrender to
the immune system which is kept in balance in reaction to the pathogen flux.

These properties characterize our model system in which we have not really
accounted adequately for the high specificity of the adaptive immune system.
With adaptation and specificity, the opposite is known to occur. An input
of bacteria, e.g. via subcutaneous injection, known historically
as the method of ``fixation abscess'' (Canuyt, 1932), may divert (for some time)
the production of antibodies which attack tissues, and as long as the
bacterial infection evolves, the immune illness is disactivated. We
should be able to take into account this effect by reintroducing the
asymmetry between the two compartments of the immune system, a task left
for a future report.

\vskip 3mm

{\bf State $D$} (death)

\vskip 2mm

The expressions of the cell numbers in the State $D$ as well as the domain
of stability found here are identical to those (\ref{fractD})  and
(\ref{stabD}) reported in Section \ref{jhkke}. The absence of normal,
infected and immune cells characterizes the death state.

\section{Existence and stability of stationary states with
proliferating ill cells ($A_2<0$)}

\subsection{Choice of parameters}

We now consider the possibility that the infected cells have a tendency to
proliferate, which can be captured by taking the coefficient
$\alpha_2$ in (\ref{f2}) to become negative. We thus impose the typical
value
\be
\label{al2}
\al_2 = -1 \; .
\ee
This is the principal difference compared with condition (\ref{al125}). For
the other decay rates, we assume the same values, as in Eq. (\ref{al125}),
that is, $\al_1 = \al_5 = 1$.
The apoptosis rate of the immune system is denoted by $\al$, as in
Eq.(\ref{al}). For all other coefficients, we take the same values as in
Eqs. (\ref{a11}), (\ref{aii}), and (\ref{aij}) and, as in the previous
section \ref{jkhkybgr}, we consider the two possibilities of absence and
presence of an auto-immune disorder.

A remark is in order to justify the choice $a_{22}=0$ for $\al_2<0$.
Recall that the choice $a_{22}=0$ was justified
for $\al_2 >0$ by the fact that the term associated with the coefficient
$a_{22}$ provided only a small second-order effect in the limitation of
the number of infected cells. For $\al_2<0$, the linear term leads in
principle to an exponential explosion and the
second order term associated with the coefficient $a_{22}$ should then become
relevant. However, our study of the influence of this term
shows that the inclusion of the term with coefficient $a_{22}$ introduces
a significant complication in the expressions of the number of cells
of the different stationary states, while keeping unchanged the
stability regions. In a nutshell, this can be understood from the fact
the other cells provide sufficient negative feedbacks on the infected
cells to maintain their number finite, so that the second-order term
proportional to $a_{22}$ remains of minor importance, and can therefore
be safely dropped out of the equations, without a significant alteration
of the results.

\subsection{Infected cells can proliferate and no auto-immune disorder}

In absence of auto-immune disorder as in Subsection 4.2, condition
(\ref{a13}) holds. Taking into account condition (\ref{al2}) for the tendency
of infected cells to proliferate,
the forces of the dynamical system  (\ref{dimeq}) are
$$
f_1 = x_1 ( 1 - x_1 - x_5) \; ,
$$
$$
f_2 = x_2( 1- x_3 - x_4 - x_5) + x_1 x_5 \; ,
$$
$$
f_3 = x_3( x_2 - x_4 + x_5 -\al) \; ,
$$
$$
f_4 = x_4(x_2 + x_3 + x_5 -\al) \; ,
$$
\be
\label{noauto}
f_5 = x_5 ( x_2 - x_3 - x_4 -1) + \vp \; .
\ee
The analysis shows that there are again four stable stationary states,
which confirms the surprising structural stability of the dynamical
system. These states and the region of their stability are presented in
Fig. 7.

\vskip 3mm

{\bf State} $A_{\rm chr}$

\vskip 2mm

The existence of a fixed point
describing stationary solutions in which infected cells exhibit the
tendency of proliferation can be interpreted as the presence of a
chronic disease. Therefore, we mark the state as $A_{\rm chr}$.
The expressions for the stationary
solutions in State $A_{\rm chr}$ are extremely cumbersome. Their numerical
analysis is illustrated in Fig. 8. Approximate expressions of the cell and
pathogen numbers for $\al<1$ and $\vp>1$ are
$$
x_1^* \; \simeq \; (1 - \al) \left ( 1 + \frac{\al^2}{\vp} + 2\;
\frac{\al^4}{\vp^2} \right ) \; ,
$$
$$
x_2^* \; \simeq \; (1-\al) \left ( \frac{\al^2}{\vp} + 2\;
\frac{\al^4}{\vp^2} \right ) \; ,
$$
$$
y^* \; \simeq \; \frac{\vp}{\al}\; - \; \al + (1-\al)
(1+2\al)\; \frac{\al}{\vp} \; + \; ( 1 - \al)(1 + 5\al) \;
\frac{\al^3}{\vp^2} \; ,
$$
$$
x_5^* \simeq \al - (1 -\al)\; \frac{\al^2}{\vp}\; - \;
2(1-\al) \; \frac{\al^4}{\vp^2}  \; .
$$
The asymptotic behavior here is analogous, up to $O(1/\vp)$ terms,
to Eqs. (\ref{ass1}). We thus have, for $\vp \to +\infty$
\be
x_1^* \to 1-\alpha~, ~~~x_2^* \to 0~,~~~
y^* \to {\vp \over \alpha}~,~~~x_5^* \to \alpha~.
\label{wrtbrtfq}
\ee
This limit is the same as (\ref{wrtbrt}) of Section 4.2.

One observes a first important difference with the situation described
in Subsection 4.2 for State $A$. In State $A_{\rm chr}$,
one can now observe large dependencies of $x_1^*$ and $x_5^*$
as a function of $\vp$ accompanied by the large growth of the number
of infected cells as $\al$ and/or $\vp$ increase.

State $A_{\rm chr}$  for $\al <1$ survives under any pathogen flux
$\vp\geq 0$. The domain of unique existence of State $A_{\rm chr}$
extends also to the region $1 < \alpha < 2 - \vp$ with $\vp <1$. There
is a third domain limited by the the vertical line $(\alpha =1; \vp \geq 1)$,
$(\alpha +\vp=2; \vp < 1)$ and the curved line shown in Figure 7 in which
bistability occurs: state $A_{\rm chr}$ can coexist with one of the
States
$B$, $C$, or $D$ in the intersections with their respective domains of
stability, which are described below.

\vskip 3mm

{\bf State} $B_{\rm can}$

\vskip 2mm

The stationary cell numbers corresponding to this state are
\be
\label{weakB}
x_1^* = 0 \; , \qquad x_2^* = 1 - \vp \; , \qquad x_3^* = x_4^* = 0 \; ,
\qquad y^* = 0 \; , \qquad x_5^* = 1 \; .
\ee
As compared to State $B$ of Subsection 4.2, defined by Eqs. (\ref{fract}),
the State  $B_{\rm can}$, described by Eqs. (\ref{weakB}), does not contain
healthy cells, but only infected cells, and pathogens, without any normal
or immune cells. The stability region of this state is defined by the
inequalities
\be
\label{stab55}
\al+\vp > 2 \; , \qquad 0 < \vp < 1 \; .
\ee

We interpret this State  $B_{\rm can}$ as a form of cancer, for
instance Cachexy, in which infected (which proxy for tumoral) cells have taken over while the
immune system does not react anymore. Note also that the number of
pathogens is large, even for very small pathogen fluxes.
The proliferation propensity of the infected cells has developed
an endogenous illness. This form of cancer can be seen as an inherent
stable organization of the organism, when the immune system is weak
and the body does not impede the spontaneous growth of infected cells.
%This is what occurs when a virus integrates into the host genome/cell to produce a viral-induced cancer, such as human papillomavirus and cervical cancer. Another possible interpretation is to view the nucleation of malignant mutations as a constant influx and ``infection'' as meaning ``malignant.''

\vskip 3mm

{\bf State} $C$ (critical illness)

\vskip 2mm

This state is identical both in the expression of the number of cells
and its domain of stability to that studied in section 4.2.

\vskip 3mm

{\bf State} $D$ (dead)

\vskip 2mm

The death state is also identical both in the expression of the number of cells
and its domain of stability to that studied in section 4.2.

\vskip 2mm
As is seen in Fig. 7,
there are three bistability regions that can be denoted as
\begin{eqnarray}
\nonumber
\left. \begin{array}{l}
A_{\rm chr}  + B_{\rm can} \\
A_{\rm chr} + C \\
A_{\rm chr} + D \end{array} \right \}
{\rm bistability \; regions.}
\end{eqnarray}
The existence of bistability means that the state of the organism
depends on the initial conditions. In other words, it is history dependent.
Consider for instance the domain of coexistence of $A_{\rm chr}$
and $B_{\rm can}$. Each of these two states is characterized by its basin
of attraction in the space of the variables $\{ x_i|\;  i=1,2,3,4,5\}$. If
the variables are initially, say, in the basin of attraction of $A_{\rm chr}$,
the cell numbers will converge with time towards the values associated
with State $A_{\rm chr}$. On the other hand,  if the variables
are initially  in the basin of attraction of $B_{\rm can}$, the
cell numbers will converge with time towards the values associated with
State $B_{\rm can}$.

Using standard results in the theory of stochastic processes, we predict
that, under the presence of stochastic forcing such as occurring if $\vp$
has a noisy component for instance, the existence of bistability implies
that the organism can spontaneously jump from one state to the other with
which it shares its domain of stability in the $(\alpha, \vp)$ plane.
Thus, an organism with chronic disease (State $A_{\rm chr}$) may
acquire ``spontaneously'' a cancer. It also possible to imagine spontaneous
remission of cancer to a chronic disease illness. This case is actually
at the basis of immuno-therapies based on the danger theory (Matzinger, 2002).

These results also
apply to the two others domains of stability with even more gloomy scenarios:
a chronic disease (State $A_{\rm chr}$) may lead suddenly
to a critically ill state (state $C$) or worse to death (State $D$).
Such abrupt transition is reminiscent of well-known cases, for instance
when banal hepatitis can be followed by death in less than 24 hours.
The actual realization of these transitions depend on the strength of the
``barrier'' separating the two coexisting states and on the amplitude
and nature of the stochastic forcing (which can occur in the pathogen flux,
as well as in variations of other characteristics of the organism). 
The study
of these transitions is left to another paper.

\subsection{Infected cells can proliferate and auto-immune disorder }

The situation here is analogous to that of Subsection 4.3, with the
important change of the sign of the decay rate of infected cells, so that
condition (\ref{al2}) now holds. The forces of the dynamical system
(\ref{dimeq}) are thus given by
$$
f_1 = x_1 ( 1 - x_1 - x_3 - x_4 - x_5) \; ,
$$
$$
f_2 = x_2 ( 1 - x_3 - x_4 - x_5) + x_1 x_5 \; ,
$$
$$
f_3 = x_3 ( x_1 + x_2 - x_4 + x_5 - \al) \; ,
$$
$$
f_4 = x_4(x_1+x_2+x_3+x_5-\al) \; ,
$$
\be
\label{auto60}
f_5 = x_5 ( x_2 - x_3 - x_4 - 1) + \vp \; .
\ee
Again there are four types of stable stationary solutions, whose domains
of stability in the plane $(\alpha, \vp)$ are presented in Fig. 9.

A common denominator of all four states is that the number  $x_1^*$ of
normal cells is vanishing, showing that, when infected cells can
proliferate and there is an auto-immune disorder, the organism is badly
ill (typical of a post liver cirrhosis involving viral infection and auto-immune
disorder which in 10\% of cases evolves into an primitive hepatocellular carcinoma).
Contrary to the case of Subsection 5.2, here there are no bistability
regions. The destiny of the organism does not depend on its initial
conditions, but is prescribed by the organism parameters and pathogen
flux. The description of the four states is as follows.

\vskip 3mm

{\bf State} $A_{\rm inf}$ (Strong infection)

\vskip 2mm

The numbers of cells are
\be
\label{sol61}
x_1^*= 0 \; , \qquad x_2^* = \frac{2\al-\al^2-\vp}{2-\al} \; ,
\qquad y^* = \frac{2-\al-\vp}{2-\al} \; , \qquad
x_5^* = \frac{\vp}{2-\al} \; .
\ee
This state is stable in the region of the $\al-\vp$ plane where
\be
\label{stab62}
 0 < \al \leq 1 \; , \qquad \vp < \al(2 -\al)
\ee
and
\be
\label{stab63}
1 < \al < 2 \; , \qquad \al + \vp < 2 \; .
\ee
Though the organism is still alive, it is very ill, since there
are no healthy cells. This constitutes a drastic difference compared with State
$A_{\rm aut}$ of Subsection 4.3. The behaviors of the normalized cell numbers and
pathogens, defined by Eqs. (\ref{sol61}), are shown in Fig. 10 and can be
qualitatively understood as the results of the additive effects of the two
defects of this organism (infected cells can proliferate and there is
an auto-immune disorder).

\vskip 3mm

{\bf State} $B_{\rm can}$
\vskip 2mm

This state is identical both in the expression of the number of cells
and its domain of stability to that studied in section 5.2.

\vskip 3mm

{\bf State } $C$ (critical ill)

\vskip 2mm

The present State $C$ is identical in its cell numbers and in its
domain of stability in the plane $(\alpha, \vp)$ to State C
of Section 4.3.  The organism has only immune cells fighting the 
pathogens. As in State $C$ of Section 4.3, the domain of stability is 
quite large, since the organism suffers attacks from both immune cells
and pathogens.

\vskip 3mm

{\bf State} $D$ (dead)

\vskip 2mm

The death state is also identical both in the expression of the number
of cells and its domain of stability to that studied in the other
sections.

\section{Discussion}

We have derived a model of immune system homeostasis, describing
complex interactions between healthy cells, infected cells, immune
system, and pathogens. The model is represented in full
generality by a five-dimensional
dynamical system, but we have considered here a simplified four dimensional
system obtained by assuming symmetric strengths of the responses
of the innate and adaptive components of the immune system.
The richness of this system provides a classification of a rather
diverse set of afflictions typified by important human diseases.
Here, we have concentrated our attention on the investigation of the
basic topological structure of the dynamical system. Analysing the
existence of stationary states and their stability, we have discovered
that the model is surprisingly structurally stable when changing
two sets of control parameters. We have specifically explored
the influence of an auto-immune disorder and of the possibility of
infected cells to proliferate. We have thus considered four cases
represented schematically in Figure 11:
\begin{enumerate}
\item no auto-immune disorder  ($a_{13}=a_{31}=a_{14}=a_{41}=0$)
and no proliferation of infected cells ($a_2>0$)
\item auto-immune disorder  ($a_{13}=a_{31}=a_{14}=a_{41}>0$)
and no proliferation of infected cells ($a_2>0$)
\item no auto-immune disorder ($a_{13}=a_{31}=a_{14}=a_{41}=0$)
and proliferation of infected cells ($a_2<0$)
\item auto-immune disorder ($a_{13}=a_{31}=a_{14}=a_{41}>0$)
and proliferation of infected cells ($a_2<0$).
\end{enumerate}
By doing this, we consider the parameters $a_2$ and $a_{13}, a_{31}, a_{14},
a_{41}$ as determined exogenously. Their variations which span the four
above regions leads to different states, as shown in figure 11. In
reality, one would like to have a more complete description in which the
dynamics of these parameters is not imposed externally but is endogenized.
Our present approach allows us to classify the different states under the
condition that the auto-immune propensity and the proliferation tendency
of infected cells are kept fixed. Making endogenous the dynamics of these
control parameters (which in this way would become more like ``order
parameters'') may give rise to new phenomena, but this is beyond the first
exploratory scope of the present paper.

An example of the possible time evolution of the parameters
$a_{13}, a_{31}, a_{14}, a_{41}$ occurs during chronic infections
characterized by a strong response of the immune system, which may eventually
evolve to some auto-immune disorder due to the strengthening synthesis
of antibodies reacting to the membranes of the infected cells deteriorating
into attacks of the membrane of normal cells. This phenomenon occurs for
instance for hepatic cirrhosis due to hepatitis C virus (HCV), in which
cirrhosis is due to the chronic inflammatory response against the liver
cells rather than the virus itself.

In each of the four cases, we find four stable stationary states whose
boundaries are determined by the values of the system parameters and in
particular by the apoptosis rate $\alpha$ and the pathogen flux $\vp$. The
transitions between the states is reminiscent of the
phase transitions in statistical systems (Yukalov and Shumovsky, 1990;
Sornette, 2006). The occurrence of a given state essentially depends
on the balance between the strength of the immune system, characterized
by its apoptosis rate, and the external influx of pathogens, which 
supports the concept that the homeostasis of the organism is
governed by a competition between endogenous and exogenous factors.
It is important to stress that stable stationary states exist only if
the apoptosis in the immune system prevails over the reproduction of
immune cells, so that the effective rate $\al$ be positive.

This study and the proposed model presented in section 3 has been
motivated by the endo-exo hypothesis discussed in section 2.
The existence of State A, which describes an organism with a ``healthy''
immune system, capable of controlling any amount of exogenous pathogen flux, 
provides a clear embodiment of this concept. The fact that the
healthy state A exists only under the  influence of a sufficiently large
pathogen flux $\vp$ suggests that health is not the absence of pathogens,
but rather a strong ability to find balance by
counteracting any pathogen attack. Furthermore, we also find that the
critically ill State C can recover to State A by increasing the strength
of the immune system (decreasing the apoptosis rate $\alpha$) but evolves
to the death State D if an attempt is made of decreasing too much the
pathogen flux. This paradoxical behavior illustrates again that pathogens
seem, in our system, to be necessary to ensure recovery and health.

We thus propose that a fundamental change in the way we should think
about health and illnesses is required. As an example, a concept such as the
basic reproductive ratio $R_0$, defined as the average 
number of secondary cases transmitted by a single infected 
individual that is placed into a fully susceptible population, is useless
in our context. Indeed, $R_0$ assumes that there is such a thing
as a non-infected state. Then, $R_0$ quantifies
the  initial rate of spread of a disease that appears and invades
the initially non-infected population. In the standard classification,
there will be an epidemic if $R_0 > 1$, while the epidemic dies out
before spreading explosively in the population if $R_0 < 1$.
In our case, $R_0$ would describe the evolution of the populations
of different cells in the organism, starting from an initial state in which $x_2 \ll 1$
(almost complete absence of infected cells), when a burst of pathogens is suddenly introduced.
As can be seen from equations (\ref{kin2}) or, in reduced form, 
(\ref{dimeq}) with (\ref{f2}), when $x_2$ is small (very few infected cells), the equation
controlling the early dynamics of infected cells is
$dx_2/dt = - x_2 \left( \alpha_2 + a_{23} y + a_{25} x_5 \right) 
+a_{51} x_1 x_5$. Since $a_{23}$ and $a_{25}$ are always positive,
when $\alpha_2>0$, 
the initial {\it proportional} growth rate (the bracket term
in factor of $x_2$ in the r.h.s.) is always positive, and minus this term is always negative, 
so that $R_0$ is by definition always
less than $1$. In the case where $\alpha_2<0$ (referred to above as
the regime with ``proliferation of infected cells), $R_0$ can be larger than $1$
if $ \alpha_2 + a_{23} y + a_{25} x_5$ is negative. However, this
analysis is misleading because it neglects the 
different channel $a_{51} x_1 x_5$, in the presence
of the complex feedback loops described by equations (\ref{dimeq}) with (\ref{f2})
and the fact that any set of parameters corresponds to a well-defined
equilibrium state (except for the case of proliferating infected cells in the absence
of auto-immune disorder, in which there is a zone of coexistence of two equilibrium
states as shown in figure 7). The argument on the uselessness of the concept of $R_0$
in our context can be made clearer
by considering an hypothetical pure organism with initial values $x_2=y=x_5=0$,
i.e., made only of normal cells, for which
$R_0 = 1-\alpha_2$. Hence, the sign of $\alpha_2$ determines
the initial spread of the disease, according to the standard approach. 
But, as our analysis has shown, this measure is completely misleading
and cannot represent the complex reality of the dynamical evolution
of the organism into an homeostatic state characterized by a subtle
balance between the immune system response to ever present
pathogens and infected cells. 
Our approach has emphasized that a correct description
of the homeostatic equilibrium of an organism is through the determination
of the stable fixed points of the system dynamics. Once the organism is
in a given stationary state, its dynamics is controlled by the Lyapunov
exponents which take automatically into account
the different coupling terms between the dynamical variables.
By definition of the stability of the fixed points, the Lyapunov exponents are all 
negative. In contrast, the concept of the basic reproductive ratio implies
that it makes sense to imagine  the situation of a completely non-infected organism which is
suddenly plunged into a situation described by a specific value
of the pair $(\alpha, \phi)$ and of the parameters controlling for
the existence or absence of an auto-immune disorder and of the 
proliferation of infected cells. The basic reproductive ratio would then
quantify how this hypothetical organism would see its number of
infected cells  grow initially, to eventually converge to the 
corresponding equilibrium state associated with these control parameters.

It is possible to generalize the concept of the basic reproductive ratio 
into the rate of evolution from one state to a neighboring one
when the border between two domains of stability is crossed. But
this corresponds to a different approach, based on changing the
control parameters, in contrast with the initial-value approach
implied by the basic reproductive ratio $R_0$. Such dynamical
evolution is left for another study.

Our results provide insights in so-called regeneration medicine
(spontaneous recovery from near extinction, or similarly defined).
Actually, our phase diagrams illustrate several cases where regeneration
can spontaneously occur.  The most obvious one is shown in figure 7 in
the coexistence regions for the case of a patient with proliferating ill
cells in the absence of auto-immune disorder. In the presence of ubiquitous
random perturbations and noise, we predicted that a patient in the
critically ill state $C$ or in the cancer state $B_{\rm can}$ can
spontaneously recover (partially) to the chronically ill state
$A_{\rm chr}$. This regeneration can be viewed as an ``emergent''
property of the immune system, in the sense that it is a fundamental
equilibrium structure derived non-trivially from the basic rules of
interactions between the five compartments of the organism.
Another paradoxical instance where spontaneous regeneration occurs
is shown in figure 1: as discussed in section 4.2, a patient with
an evanescent and highly vulnerable immune system may recover to the
healthy state $A$ upon an {\it increase} of the flux of pathogens,
all other things remaining equal! By making endogenous the dynamics
of $\alpha$ and $\phi$, no doubt that other paradoxical and surprising
regeneration processes will be found. In light of the increasing
importance of regenerative medicine and the promise it holds for the
future of the critically ill and injured, our model provides an attractive
starting point to formulate the different instances
of regeneration that can exist.

It goes without saying that, in real life, the parameters of the
dynamical system, characterizing any organism, are not necessarily
constant, and external conditions are always varying. Therefore,
the next step requires us to study the dynamics of the suggested model,
including random variations of the pathogen influx and of
system parameters. It would, however, be unreasonable to blindly fix
the parameters, having so many of them. We plan in our future work
to consider the model dynamics specifying the parameters for some
particular medical situations.

\newpage

\newpage

{\bf{\Large Appendix A. Lyapunov Stability Analysis}}

\vskip 5mm

The stability analysis of the considered system of equations is based
on the following Lyapunov theorem (see, e.g., Chetaev 1990).

\vskip 2mm

{\bf Theorem}. {\it Let us consider a system of nonlinear autonomous
differential equations
$$
\frac{dx_i}{dt} = f_i(x), \qquad \qquad \qquad \qquad (A.1)
$$
where $x\equiv\{ x_i: \; i=1,2,\ldots,n\}$. Let $x^*=\{ x_i^*\}$ be a
set of stationary solutions to Eq. (A.1) and the functions $f_i(x)$,
for all $i=1,2,\ldots,n$, be holomorphic in the vicinity of $x^*$. And
let $J_i(x)$ be the eigenvalues of the Jacobian matrix
$[J_{ij}(x)]=[\prt f_i/\prt x_j]$.

If all Lyapunov exponents $\lbd_i\equiv{\rm Re}J_j(x^*)$ are negative,
then the solution $x^*$ is asymptotically stable, hence, Lyapunov stable.
In contrast, if at least one of the Lyapunov exponents $\lbd_i$ is positive, 
then the solution $x^*$ is unstable.}

\vskip 2mm

In our case, the right-hand sides of Eq. (\ref{dimeq}), defined in Eqs.
(\ref{f1}) to (\ref{f5}), are holomorphic by construction. Thus, the 
condition for the validity of the Lyapunov theorem is satisfied. We 
accomplished the stability analysis for each of the four different 
situations investigated in our paper.

\vskip 2mm

(i) For the case of decaying ill cells without auto-immune disorder, 
the dynamical system is characterized by the right-hand sides of Eqs. 
(\ref{dineq1}) and (\ref{y}). The equations for the stationary 
solutions are
$$
x_1( 1 - x_1 - x_5) =0 \; , \qquad
x_2 ( 1 + x_5 +y) - x_1x_5 = 0 \; ,
$$
$$
y( x_2 + x_5 - \al) =0 \; , \qquad
x_5 ( x_2 - y -1 ) +\vp = 0 \; .
$$
This system of equations has 10 sets of stationary solutions 
$\{ x_1^{(k)},x_2^{(k)},y^{(k)},x_5^{(k)}\}$, with $k=1,\ldots,10$.

Out of these 10 fixed points, only four of them are admissible
as possible states of the organism, in the sense that all their normalized
cell concentrations are non-negative. The other 6 fixed points
are not considered further, as they lie in the non-biological part
of the phase diagram $\{ x_1, x_2, y, x_5\}$. The four admissible
fixed points are denoted  $A$, $B$, $C$, $D$ (see classification
of these solutions and their expressions in Sec. 4.2). The corresponding
domain of existence of these four different fixed points are presented in Fig. 1.

Then, following closely the Lyapunov theorem, we calculate the Jacobians
for each of the four sets $A$, $B$, $C$, $D$ of admissible fixed points and determine the
corresponding eigenvalues. We check if the real parts of the eigenvalues are negative.
The procedure is straightforward. The analytical expressions for the 
eigenvalues were obtained by Mathematica. We do not provide here the exact 
formulas of the eigenvalues because they are extremely cumbersome. Numerical, 
and when possible, analytical investigations of the real parts 
of the eigenvalues as functions of parameters $\al$ and $\vp$ have been 
performed. These investigations led to the determination of the 
regions of stability (negative values of the real parts of the eigenvalues)
for each of the four fixed points according to the phase diagram
shown in Fig. 1.

\vskip 2mm

(ii) The case of decaying ill cells with auto-immune disorder is described by
the functions (\ref{eqII}). The equations for the stationary solutions are
$$
x_1 ( 1 - x_1 - x_5 - y) = 0\; , \qquad
x_2 ( 1 + x_5 + y) -  x_1 x_5 = 0\; ,
$$
$$
y ( x_1 + x_2  + x_5 -\al) = 0 \; , \qquad
x_5 ( x_2 - y -1) + \varphi = 0 \; .
$$
There are 9 sets of solutions to this system of equations, but only 4 of them
are admissible, having all non-negative normalized cell concentrations.
These 4 sets of solutions 
are again denoted as $A$, $B$, $C$, $D$, and are presented in Sec.4.3. 

The stability of these four stationary solutions is checked by following the
Lyapunov theorem. Our investigations show that all real parts of the
eigenvalues of the corresponding Jacobian matrices are negative inside the
regions shown in Fig. 5, qualifying the stability of these fixed points
$A$, $B$, $C$, $D$ in their respective domains of existence.

\vskip 2mm

(iii) The homeostasis with proliferating ill cells without auto-immune
disorder is characterized by the functions given by (\ref{noauto}). The stationary
solutions are given by the equations
$$
x_1 ( 1 - x_1 - x_5) = 0 \; , \qquad
x_2 ( 1 - x_5 - y) + x_1 x_5 = 0 \; ,
$$
$$
y ( x_2 + x_5 - \al) = 0 \; , \qquad
x_5 ( x_2 - y - 1 ) + \vp = 0 \; .
$$
There are 10 sets of solutions to this system of equations, and again only 4 sets
are admissible. These sets of solutions correspond to the states $A$, $B$, 
$C$, $D$, and are presented in Sec.5.2. The regions of existence of these solutions
are shown in Fig. 7. According to the Lyapunov stability analysis, the given 
states are found to be stable in the corresponding existence regions.

\vskip 2mm

(iv) The case of proliferating ill cells with auto-immune disorder is
characterized by the dynamical system with the functions given in (\ref{auto60}).
The system of equations for the stationary solutions reads as
$$
x_1 ( x_1 + x_5 + y -1 ) = 0 \; , \qquad
x_2 ( 1 - x_5 - y ) + x_1 x_5 = 0 \; ,
$$
$$
y ( x_1 + x_2 + x_5 - \al) = 0\; , \qquad
x_5 ( x_2 - y - 1 ) + \vp = 0 \; .
$$
There are 9 sets of solutions to this system of equations. And again, only
4 sets of solutions are admissible. These sets of solutions $A$, $B$, $C$,
$D$ are described in Sec.5.3. The regions of existence of these solutions are 
shown in Fig. 9. According to the Lyapunov stability analysis, the given 
states are stable in the corresponding existence regions.

\vskip 2mm

For each of these cases, we found all stationary solutions and analysed
their stability according to the Lyapunov theorem. It turned out that
all solutions in the considered cases
(Sections 4.2, 4.3, 5.2, and 5.3) are stable in 
their domains of existence. In addition to the Lyapunov
stability analysis, we have checked the stability of the fixed points by
intensive direct numerical calculations solving the system (\ref{dimeq})
for different initial values taken in the vicinity of the stationary points
and for different parameters $\al$ and $\vp$. The direct numerical 
calculations for the whole dynamical system were found to be in complete 
agreement with the stability analysis based on the Lyapunov theorem.

\newpage

{\Large{\bf References}}
\vskip 5mm

{\parindent=0pt

Anderson, P.W., 1972.  More is Different, Science 177 (4047), 393-396.

\vskip 2mm
Arneodo, A., Coullet, P., Tresser, C., 1980. Occurrence of strange
attractors in three-dimensional Volterra equations. Phys. Lett. A 79,
259-263.

\vskip 2mm
Beisswenger C, Kandler K, Hess C, Garn H, Felgentreff K, Wegmann M,
Renz H, Vogelmeier C, Bals R. (2006)
Allergic airway inflammation inhibits pulmonary antibacterial
host defense, J Immunol. 177, 1833-1837.

\vskip 2mm
Benton, T.G., 2006. Revealing the ghost in the machine: Using spectral
analysis to understand the influence of noise on population dynamics.
Proc. Natl. Acad. Sci. 103, 18387-18388.

\vskip 2mm
Blaser, MJ. and D. Kirschner, 2007. The equilibria that allow bacterial
persistence in human hosts, Nature  449, 843-849.

\vskip 2mm
Bollinger, R.R., A.S. Barbas, E.L. Bush, S.S. Lin and W. Parker, 2007.
Biofilms in the large bowel suggest an apparent function of the human 
vermiform appendix, J. Theor. Biology, in press, 
doi:10.1016/j.jtbi.2007.08.032

\vskip 2mm
Brown, K.S., Hill, C.C., Calero, G.A., Myers, C.R., Lee, K.H.,
Sethna, J.P., Cerione, R.A., 2004. The statistical mechanics of
complex signaling networks: nerve growth factor signaling. Phys.
Biol. 1, 184-195.

\vskip 2mm
Brown, K.S., Sethna, J.P., 2003. Statistical mechanics approaches to
models with many poorly known parameters. Phys. Rev. E 68, 021904-9.

\vskip 2mm
Canuyt, G., 1932. Fixation Abscess, The Journal of Laryngology \& Otology 47, 235-242. 

\vskip 2mm
Chetaev N.G., 1990. Stability of Motion. Nauka, Moscow.

\vskip2mm
Ginoux, J.M., Rossetto, B., Jamet, J.L., 2005. Chaos in a
three-dimensional Volterra-Gause model of predator-prey type. Int. J.
Bifurc. Chaos 15, 1689-1708.

\vskip 2mm
Hofbauer, J., Sigmund, K., 2002. Evolutionary Games
and Population Dynamics. Cambridge University, Cambridge.

\vskip 2mm
Hsu, S.-B. T.-W. Hwang, Y. Kuang, 2001. Rich dynamics of a ratio-dependent
one-prey two-predators model, J. Math. Biol. 43, 377-396.

\vskip 2mm
Krammer, P.H., 2000. CD95's deadly mission in the immune system.
Nature 407, 789-795.

\vskip 2mm
Louzoun, Y., 2007. The evolution of mathematical immunology,
Immunological Reviews, 216, 9-20.

\vskip 2mm
Marsland, B.J,. Nembrini, C., Schmitz, N., Abel, B., Krautwald, S.,
Bachmann, and M.F., Kopf, M., 2005a. Innate signals compensate for
the absence of PKC-{theta} during in vivo CD8(+) T cell effector and
memory responses, Proc Natl Acad Sci U S A. 102(40), 14374-14379.

\vskip 2mm
Marsland, B.J., Battig, P., Bauer, M., Ruedl, C., Lassing, U., Beerli,
R.R., Dietmeier, K., Ivanova, L., Pfister, T., Vogt, L., Nakano, H.,
Nembrini, C., Saudan, P., Kopf, M., and Bachmann, M.F. 2005b. CCL19 and
CCL21 induce a potent proinflammatory differentiation program in licensed
dendritic cells. Immunity 22(4), 493-505.

\vskip 2mm
Matzinger, P., 2002.
The Danger Model: A renewed sense of self, Science 296, 301-305.

\vskip 2mm
Mazmanian, S.K., Liu, C.H., Tzianabos, A.O., Kasper, D.L., 2005. An
immunomodulatory molecule of symbiotic bacteria directs maturation of
the host immune system. Cell 122, 107-118.

\vskip 2mm
McFall-Ngai, M., 2007. Care for the community, Nature  445, 153-153.

\vskip 2mm
Nelson, P.W., Perelson, A.S., 2002. Mathematical analysis of delay
differential equation models of HIV-1 infection. Math. Biosciences
179, 73-94.

\vskip 2mm
Nowak, M.A., 2006.
Evolutionary Dynamics: Exploring the Equations of Life, Belknap Press.

\vskip 2mm
Palmer, C., Bick, E.M., DiGiulio, D.B., Relman, D.A., Brown, P.O.,
2007. Development of the human infant intestinal microbiota. Plos
Biology 5, 1556-1573.

\vskip 2mm
Perelson, A.S., 2002. Modelling viral and immune system dynamics.
Nature 2, 28-36.

\vskip 2mm
Perelson, A.S., Weisbuch, G., 1997. Immunology for physicists.
Rev. Mod. Phys. 69, 1219-1268.

\vskip 2mm
Reuman, D.C., Desharnais, R.A., Constantino, R.F., Ahmad, O.S.,
Cohen, J.E., 2006. Power spectra reveal the influence of
stochasticity on nonlinear population dynamics. Proc. Natl. Acad.
Sci. 103, 18860-18865.

\vskip 2mm
Schaub, B., R. Lauener, and E. von Mutius, 2006.
The many faces of the hygiene hypothesis
J. Allergy Clin. Immunol. 117(5), 969-977.

\vskip 2mm
Scott, A., 2005. Encyclopedia of Nonlinear Science. Routledge,
New York.

\vskip 2mm
Sornette, D., 2005. Endogenous versus exogenous
origins of crises. In: Extreme Events in Nature and Society, Albeverio,
S., Jentsch, V., Kantz, H. eds., Springer, Heidelberg.

\vskip 2mm
Sornette, D., 2006. Critical Phenomena in Natural Sciences. Springer,
Berlin.

\vskip 2mm
Sornette, D., Deschatres, F., Gilbert, T., Ageon, Y.,
2004. Endogenous versus exogenous shocks in complex networks: an
empirical test using book sale ranking. Phys. Rev. Lett. 93,
228701.

\vskip 2mm
Sornette, D., Helmstetter, A., 2003. Endogenous versus
exogenous shocks in systems with memory. Physica A 318, 577-591.

\vskip 2mm
Sornette, D., Malevergne, Y., Muzy, J.F., 2003. Tail risk: what
causes crashes. Risk 16, 67-71.

\vskip 2mm
Strachan, D.P.  1989. Hay fever, hygiene, and household size,
British Medical J. 299, 1259-1260.

\vskip 2mm
Xiang, Z. and X. Song, 2006. Extinction and permanence of a two-prey
two-predator system with impulsive on the predator,
Chaos Solit. Fract. 29, 1121-1136.

\vskip 2mm
Young, L.R., 1999. Artificial gravity considerations for a
Mars Exploration Mission,
Ann. NY Acad. Sci. 871, 367-378.

\vskip 2mm
Yukalov V.I., Shumovsky, A.S., 1990. Lectures on Phase Transitions.
World Scientific, Singapore.
}

\newpage

%FIGURE 1
\begin{figure}[ht]
\center
\includegraphics[width=14cm]{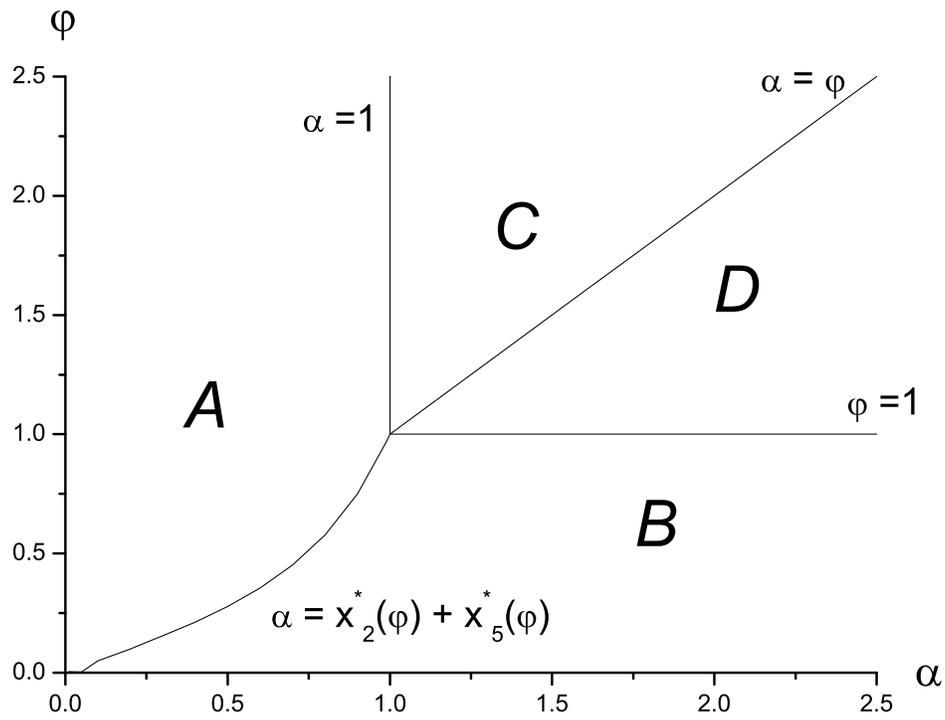}
\caption{Phase portrait in the plane $\al,\vp$ showing the stability
regions for four stationary states (A, B,C, and D), when the immune
system does not attack healthy cells (no auto-immune disorder).}
\label{fig:Fig.1}
\end{figure}

\newpage

%FIGURE 2
\begin{figure}[hbtp]
\vspace{9pt}
\centerline{
\hbox{
\includegraphics[width=7cm]{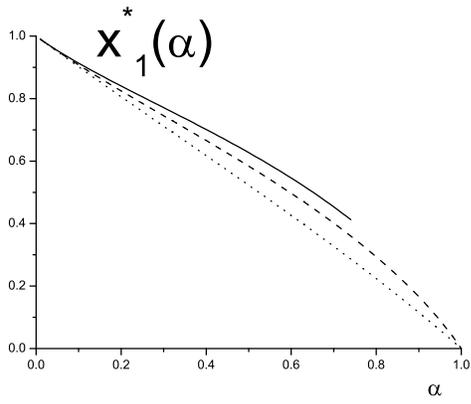} \hspace{2cm}
\includegraphics[width=7cm]{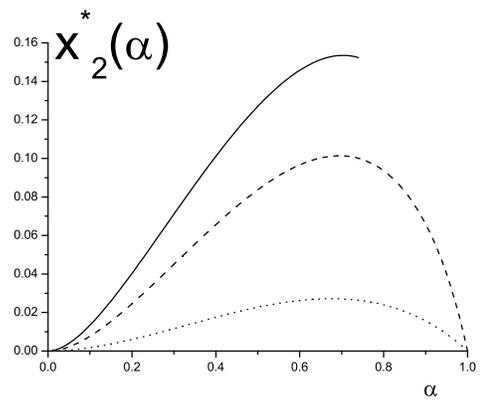} }}
\vspace{9pt}
\centerline{
\hbox{
\includegraphics[width=7cm]{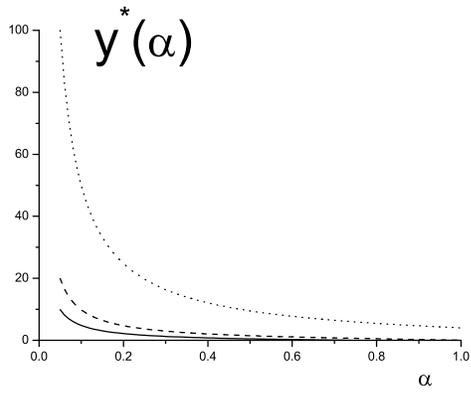} \hspace{2cm}
\includegraphics[width=7cm]{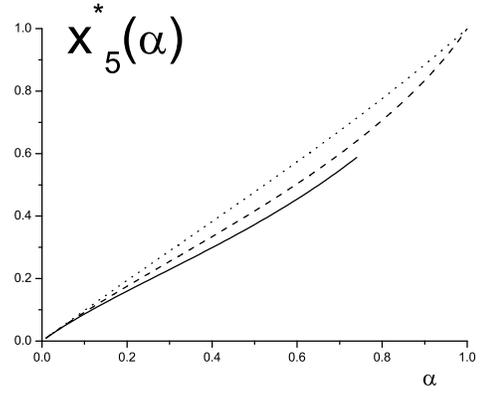} }}
\caption{State $A$: Stationary solutions for the normalized cell numbers and
pathogens as functions of the apoptosis rate $\al$ for different pathogen
fluxes $\vp=0.5$ (solid line), $\vp=1$ (dashed line), and $\vp=5$ (dotted
line) in the case of no auto-immune disorder.}
\label{fig:Fig.2}
\end{figure}

\newpage

%FIGURE 3
\begin{figure}[ht]
\centerline{
\includegraphics[width=14cm]{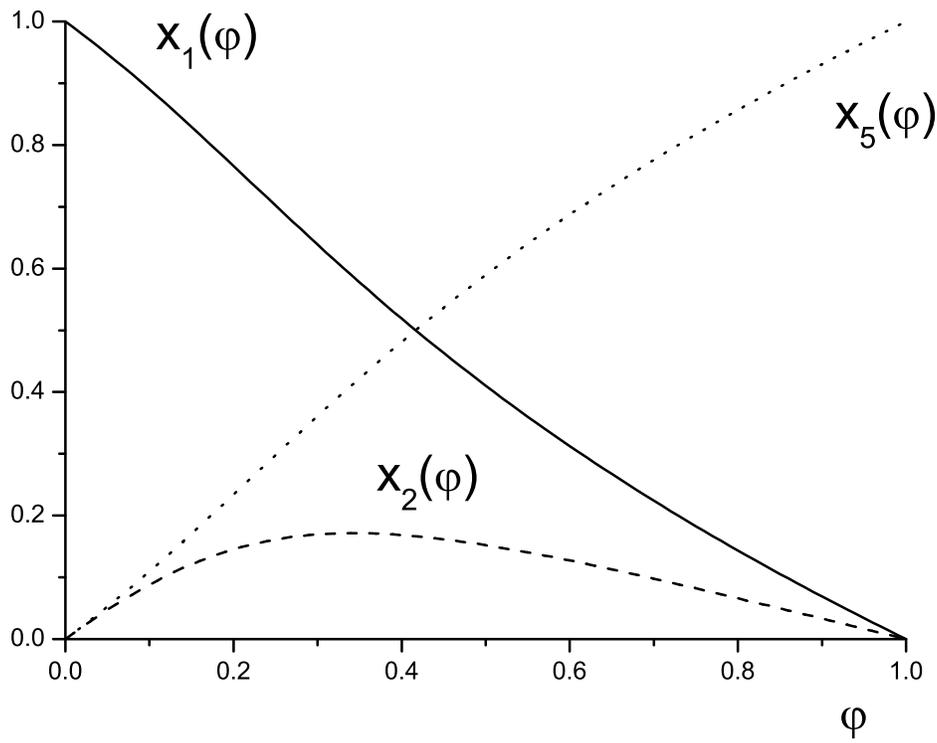}}
\caption{State $B$: Stationary solutions (\ref{fract}) for the normalized cell numbers and pathogens as functions of $\vp$ in the absence of auto-immune disorder.
$x_1(\vp)$ (solid line),
$x_2(\vp)$ (dashed line), and $x_5(\vp)$ (dotted line).}
\label{fig:Fig.3}
\end{figure}

\newpage

%FIGURE 4
\begin{figure}[ht]
\centerline{\includegraphics[width=14cm]{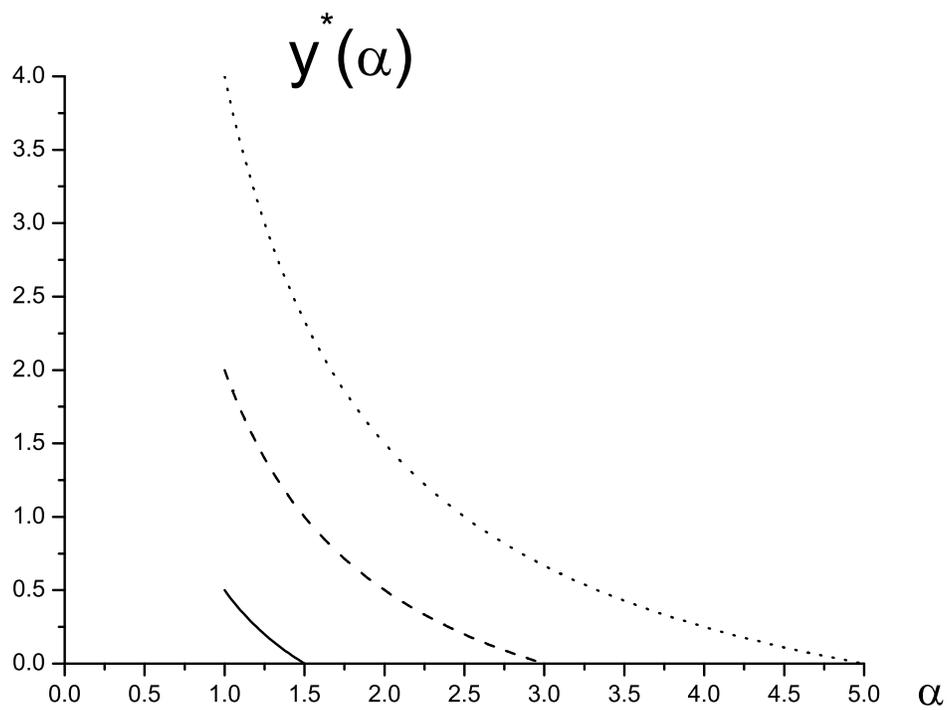}}
\caption{State $C$: Stationary solution (\ref{fractC}) for the immune
normalized cell numbers $y^*(\al,\vp)$ as a function of $\al$ for different
$\vp=1.5$ (solid line), $\vp=3$ (dashed line), and $\vp=5$
(dotted line) in the absence of auto-immune disorder.}
\label{fig:Fig.4}
\end{figure}

\newpage

%FIGURE 5
\begin{figure}[ht]
\centerline{\includegraphics[width=12cm]{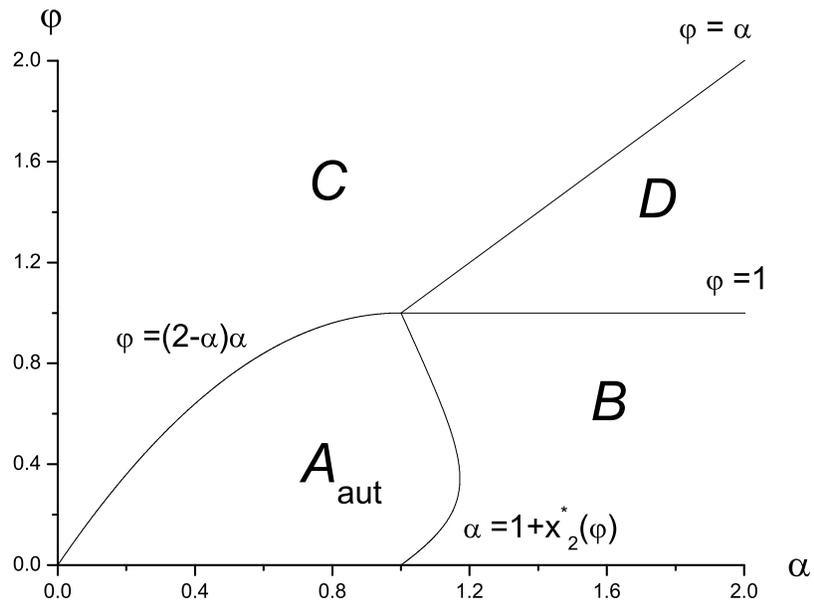}}
\caption{Phase portrait for the stability regions of the stationary
states, in the case with auto-immune disorder.}
\label{fig:Fig.5}
\end{figure}

\newpage

%FIGURE 6
\begin{figure}[hbtp]
\vspace{9pt}
\centerline{
\hbox{ \includegraphics[width=7cm]{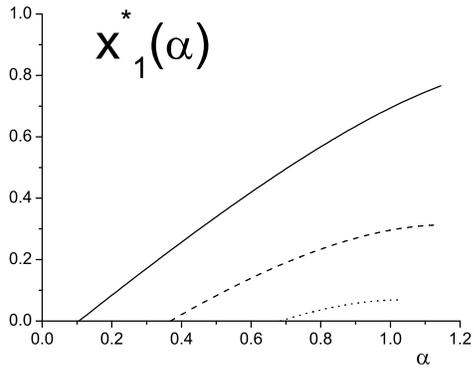} \hspace{2cm}
\includegraphics[width=7cm]{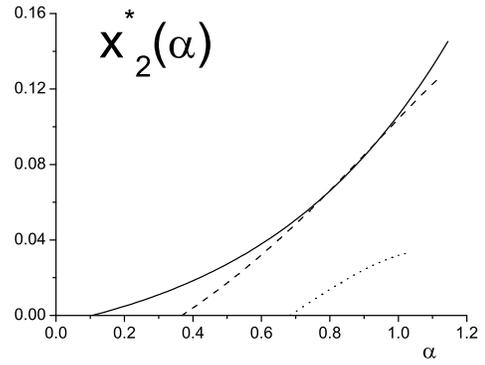} } }
\vspace{9pt}
\centerline{
\hbox{ \includegraphics[width=7cm]{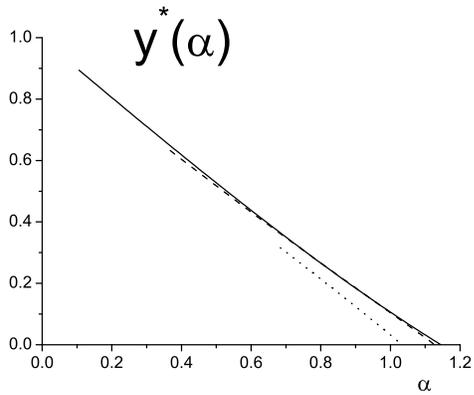} \hspace{2cm}
\includegraphics[width=7cm]{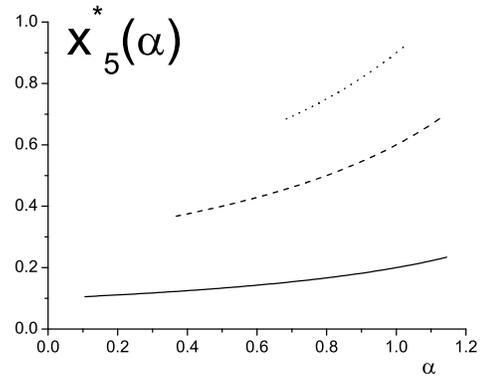} } }
\caption{State $A_{\rm aut}$: Stationary solutions for the
normalized cell numbers and pathogens as functions of the apoptosis rate $\al$ for
different pathogen fluxes $\vp=0.2$ (solid line), $\vp=0.6$ (dashed line),
and $\vp=0.9$ (dotted line) in the case with auto-immune disorder.
The solutions are shown in the region of their existence.}
\label{fig:Fig.6}
\end{figure}

\newpage

%FIGURE 7
\begin{figure}[ht]
\centerline{\includegraphics[width=14cm]{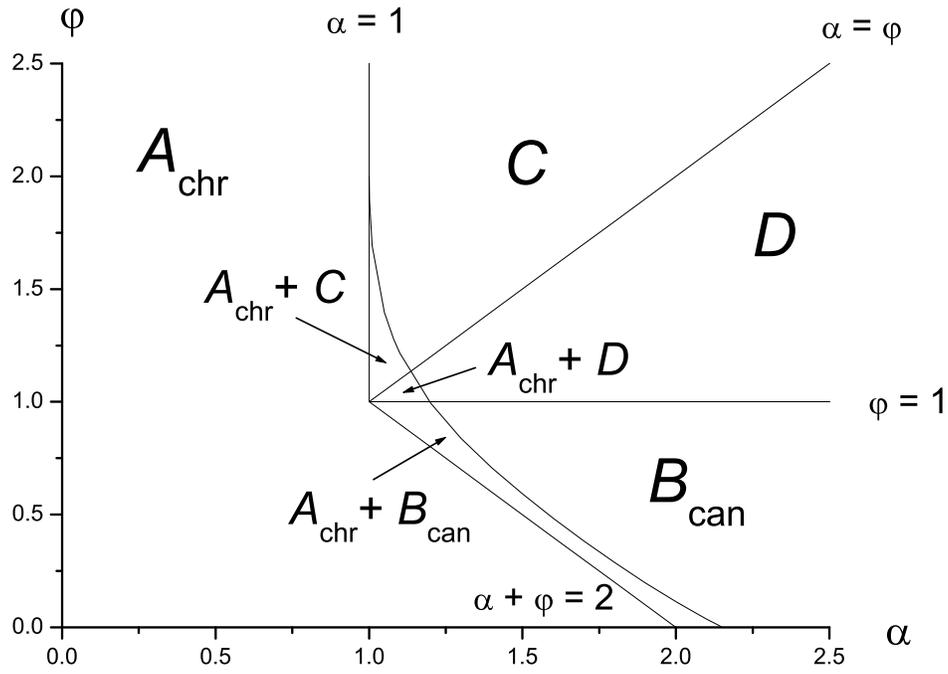}}
\caption{Phase portrait on the plane $\al,\vp$ for the case of
proliferating ill cells in the absence of auto-immune disorder}
\label{fig:Fig.7}
\end{figure}

\newpage

%FIGURE 8
\begin{figure}[hbtp]
\vspace{9pt}
\centerline{
\hbox{ \includegraphics[width=7cm]{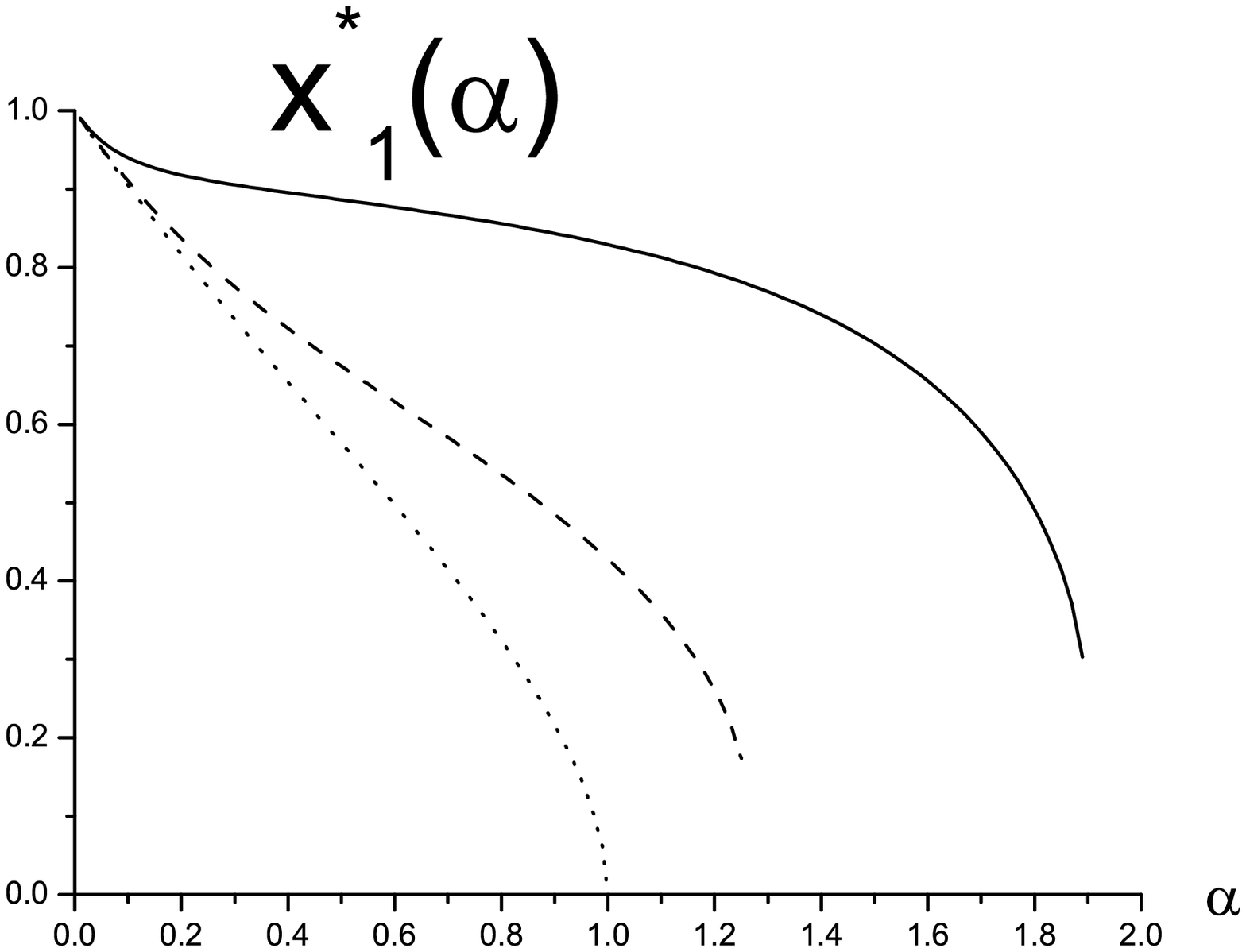} \hspace{2cm}
\includegraphics[width=7cm]{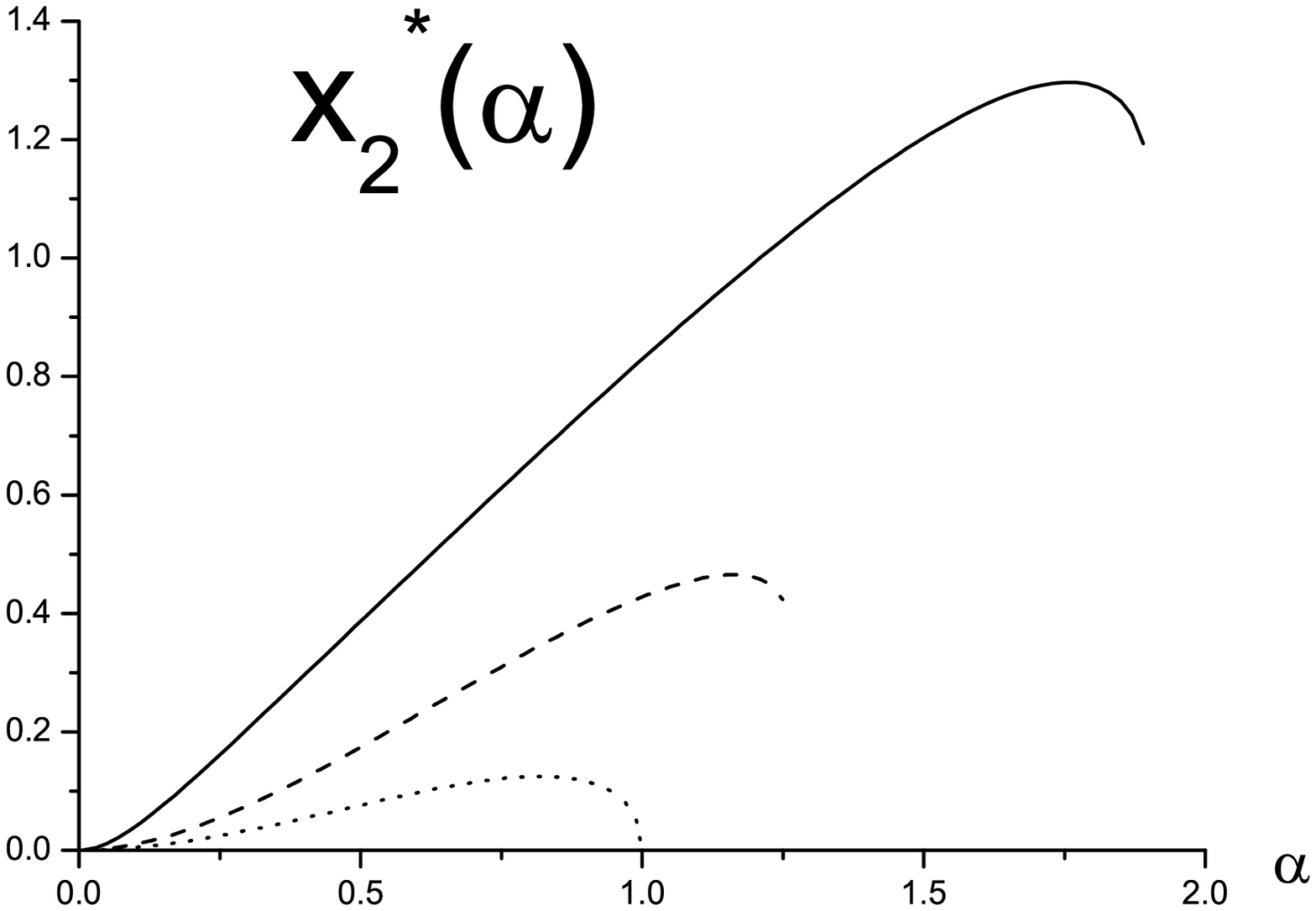} } }
\vspace{9pt}
\centerline{
\hbox{ \includegraphics[width=7cm]{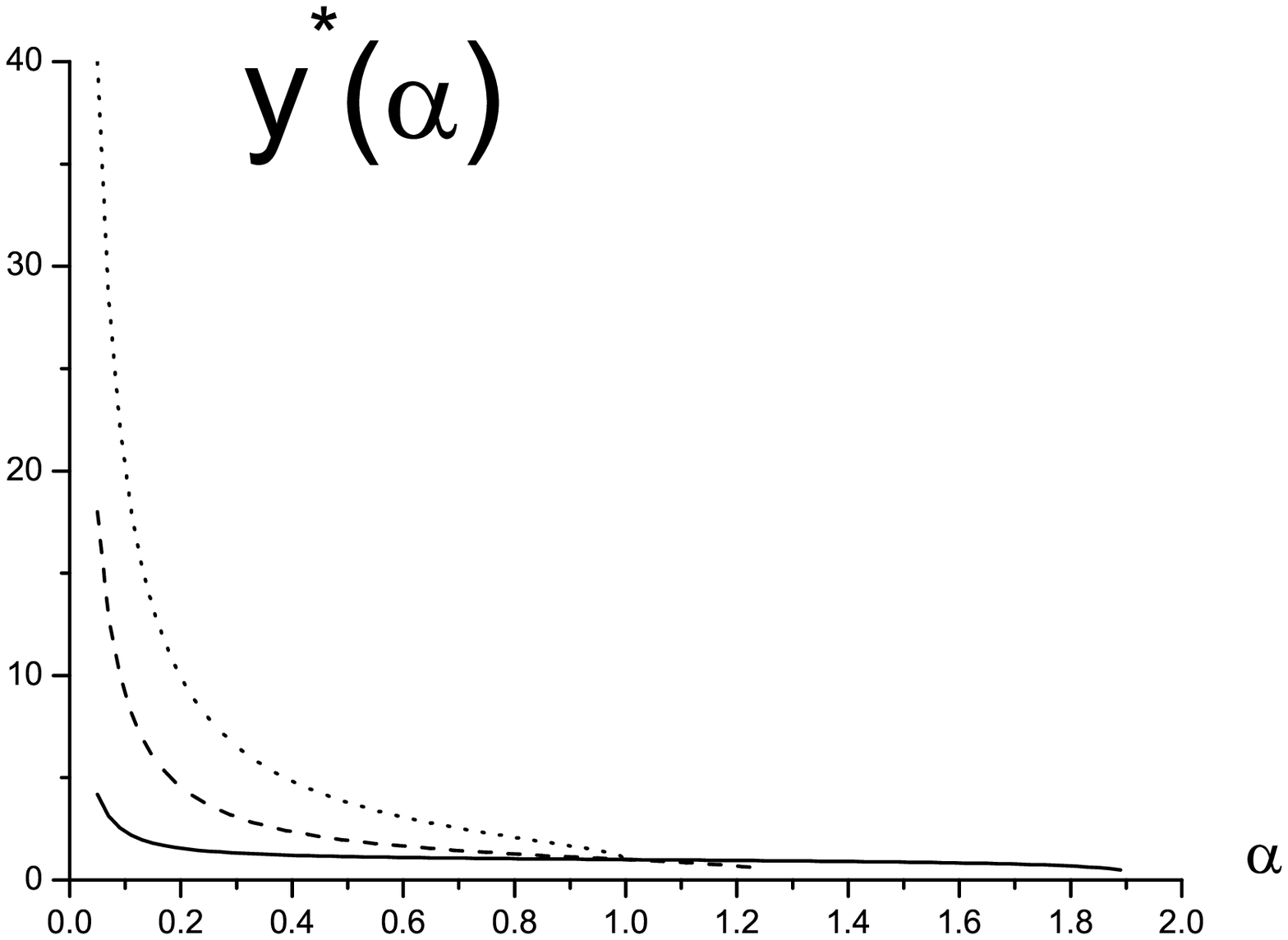} \hspace{2cm}
\includegraphics[width=7cm]{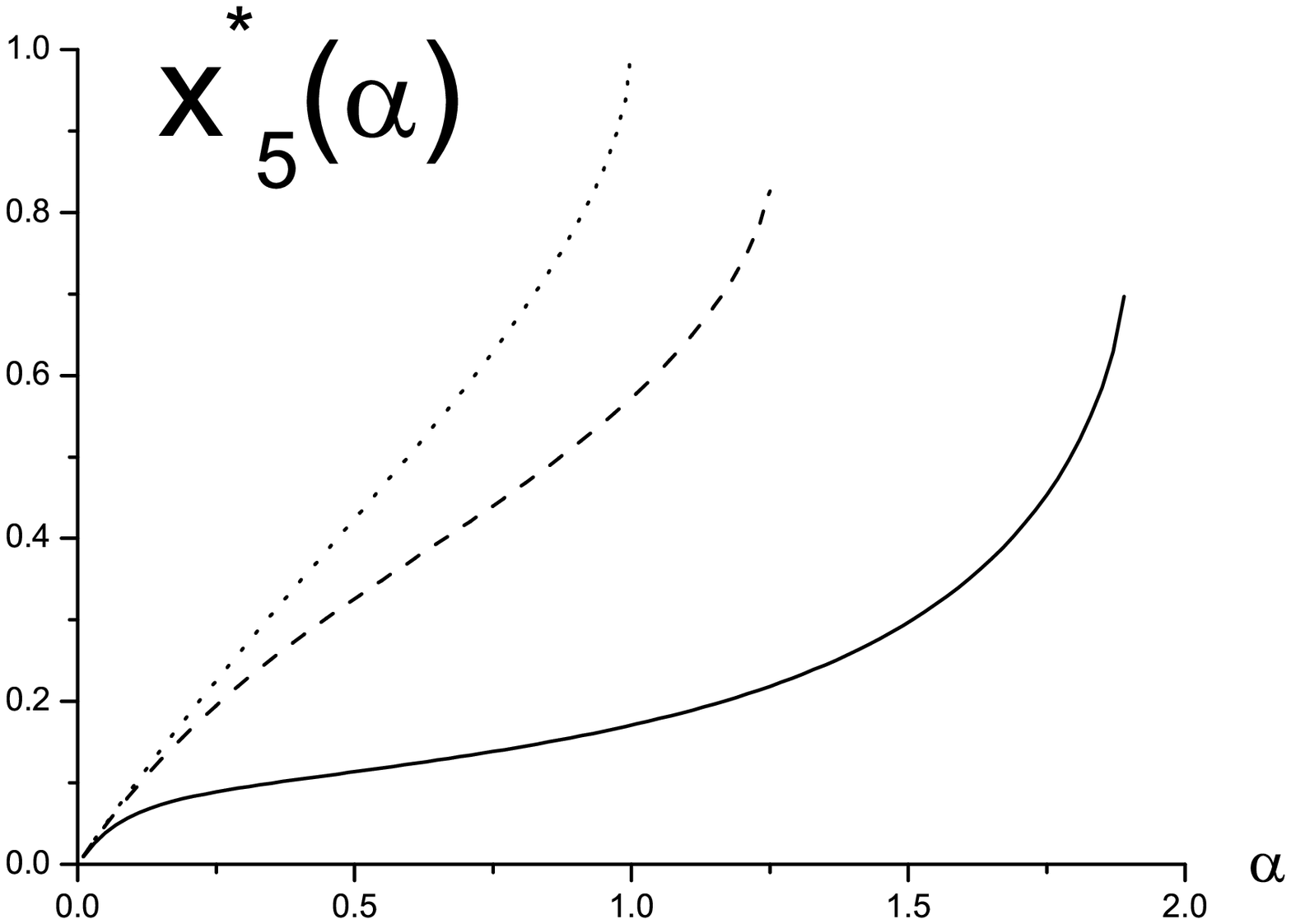} } }
\caption{State $A_{\rm chr}$: Stationary solutions for the normalized cell numbers
and pathogens as functions of the apoptosis rate $\al$ for several pathogen
fluxes $\vp=0.2$ (solid line), $\vp=0.9$ (dashed line), and $\vp=2$
(dotted line), in the case 5.2 of proliferating ill cells, in the absence of
auto-immune disorder.}
\label{fig:Fig.8}
\end{figure}

\newpage

%FIGURE 9
\begin{figure}[ht]
\centerline{\includegraphics[width=14cm]{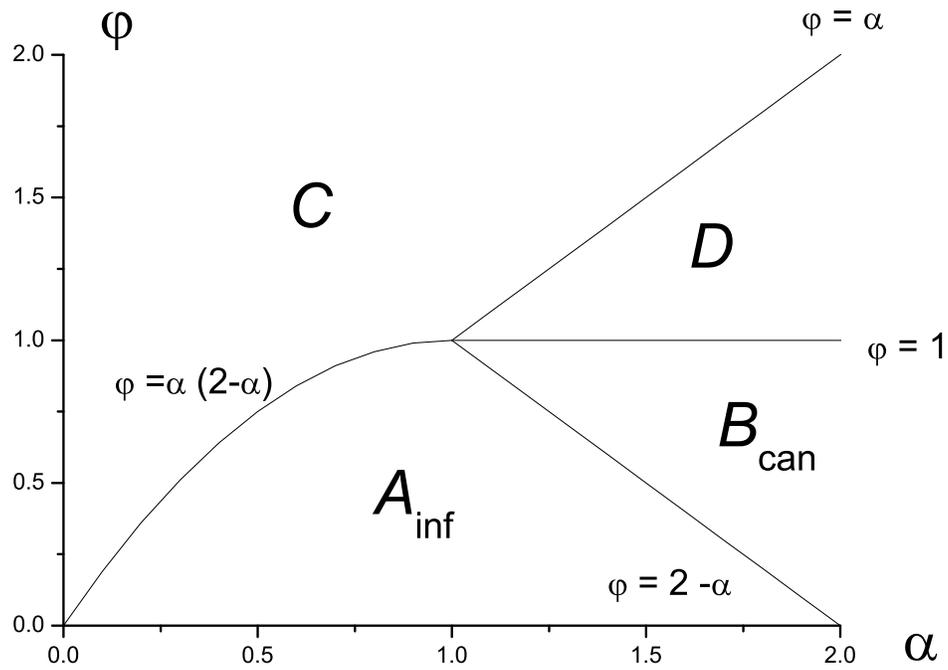}}
\caption{Phase portrait on the $\al-\vp$ plane for the stability regions
of the stationary solutions in the case of proliferating ill cells and of
auto-immune disorder.}
\label{fig:Fig.9}
\end{figure}

\newpage

%FIGURE 10
\begin{figure}[hbtp]
\vspace{9pt}
\centerline{
\hbox{ \includegraphics[width=7cm]{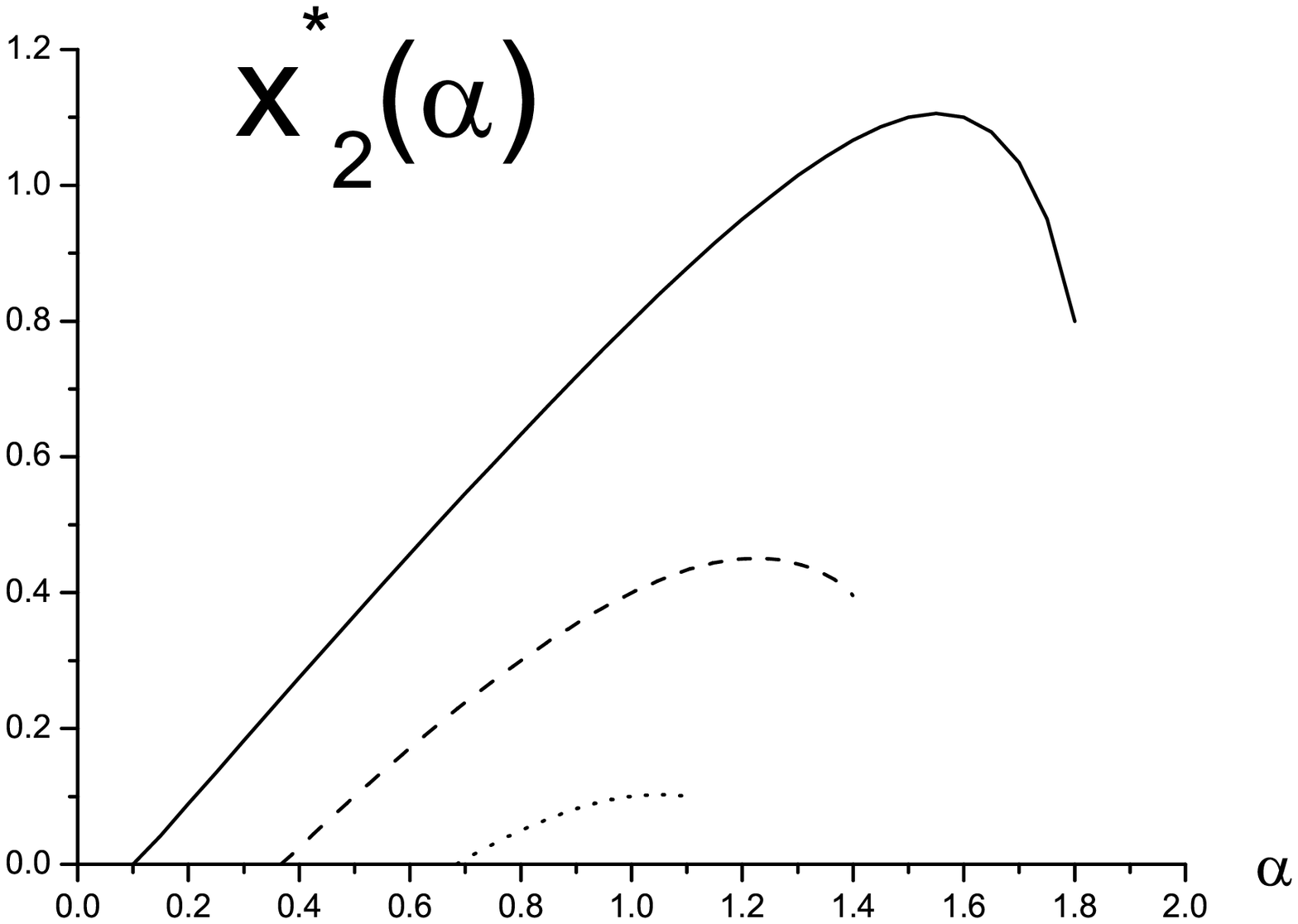} } }
\vspace{9pt}
\centerline{
\hbox{ \includegraphics[width=7cm]{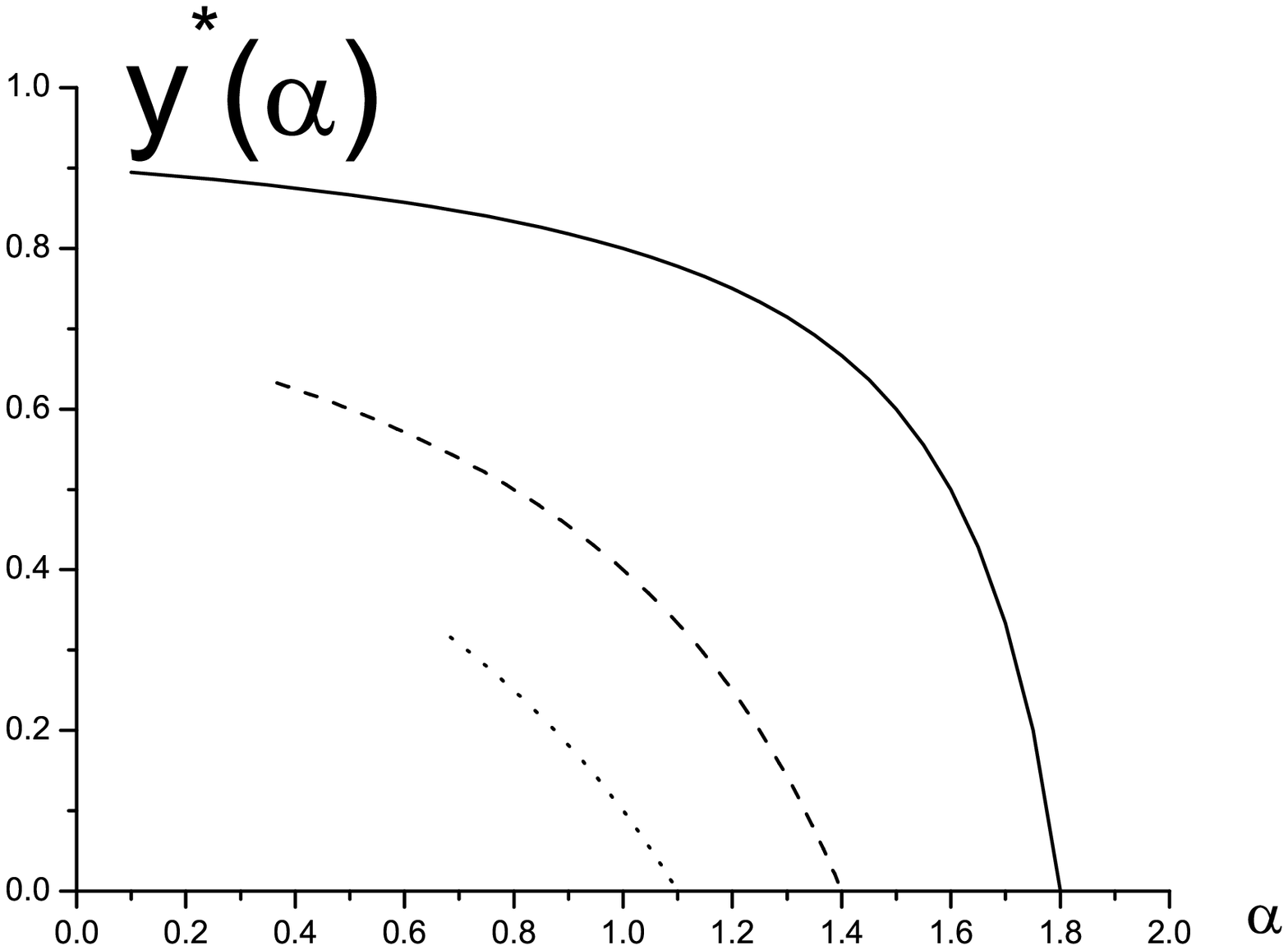} \hspace{2cm}
\includegraphics[width=7cm]{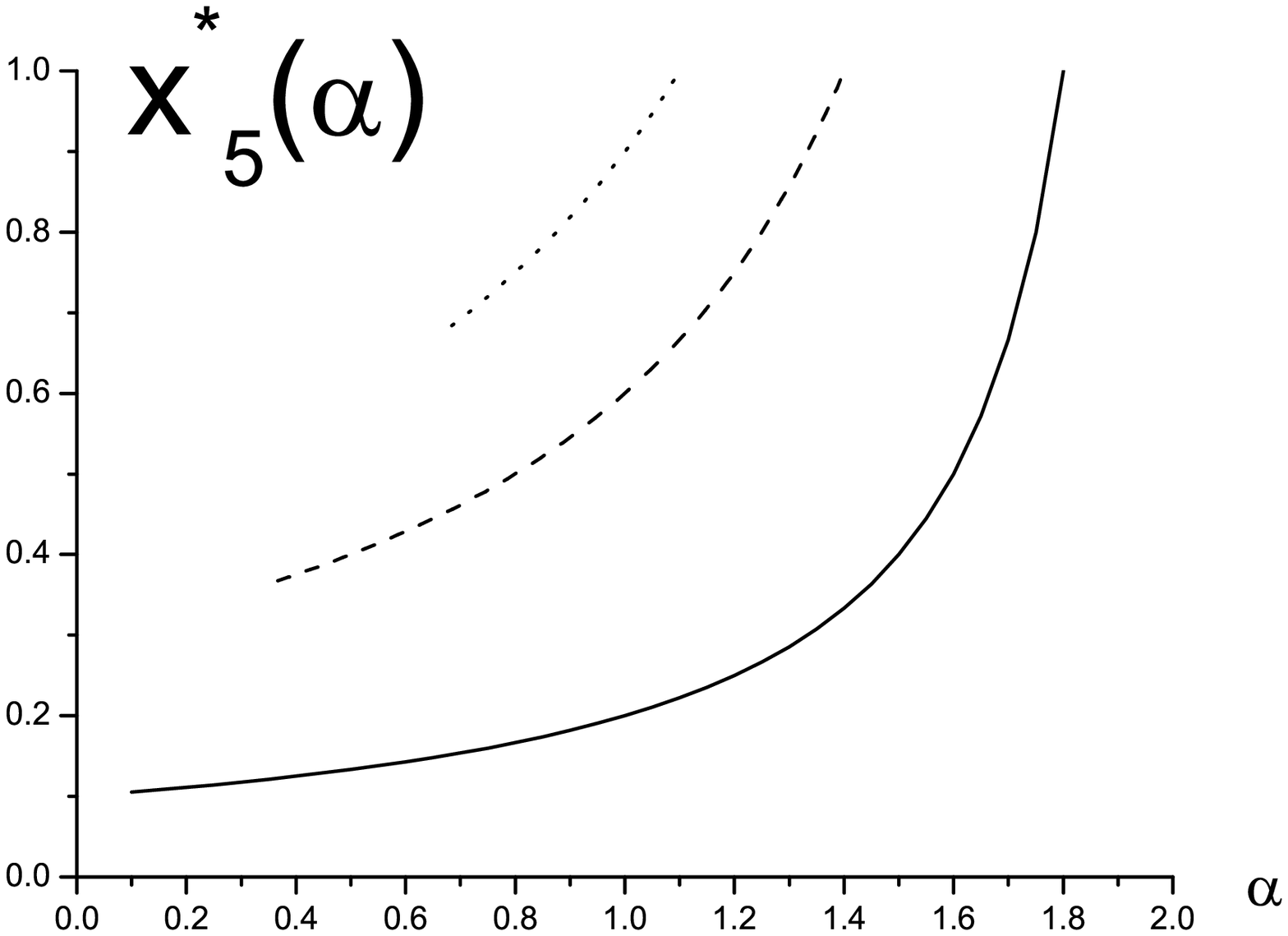} } }
\caption{State $A_{\rm inf}$ of Subsection 5.3, defined by Eqs. (\ref{sol61}).
The nontrivial solutions for the normalized cell numbers and pathogens as
functions of $\al$ for different $\vp=0.2$ (solid line), $\vp=0.6$
(dashed line), and $\vp=0.9$ (dotted line), in the case of
proliferating ill cells in the presence of auto-immune disorder.}
\label{fig:Fig.10}
\end{figure}

\newpage

%FIGURE 11
\begin{figure}[ht]
\centerline{\includegraphics[width=14cm]{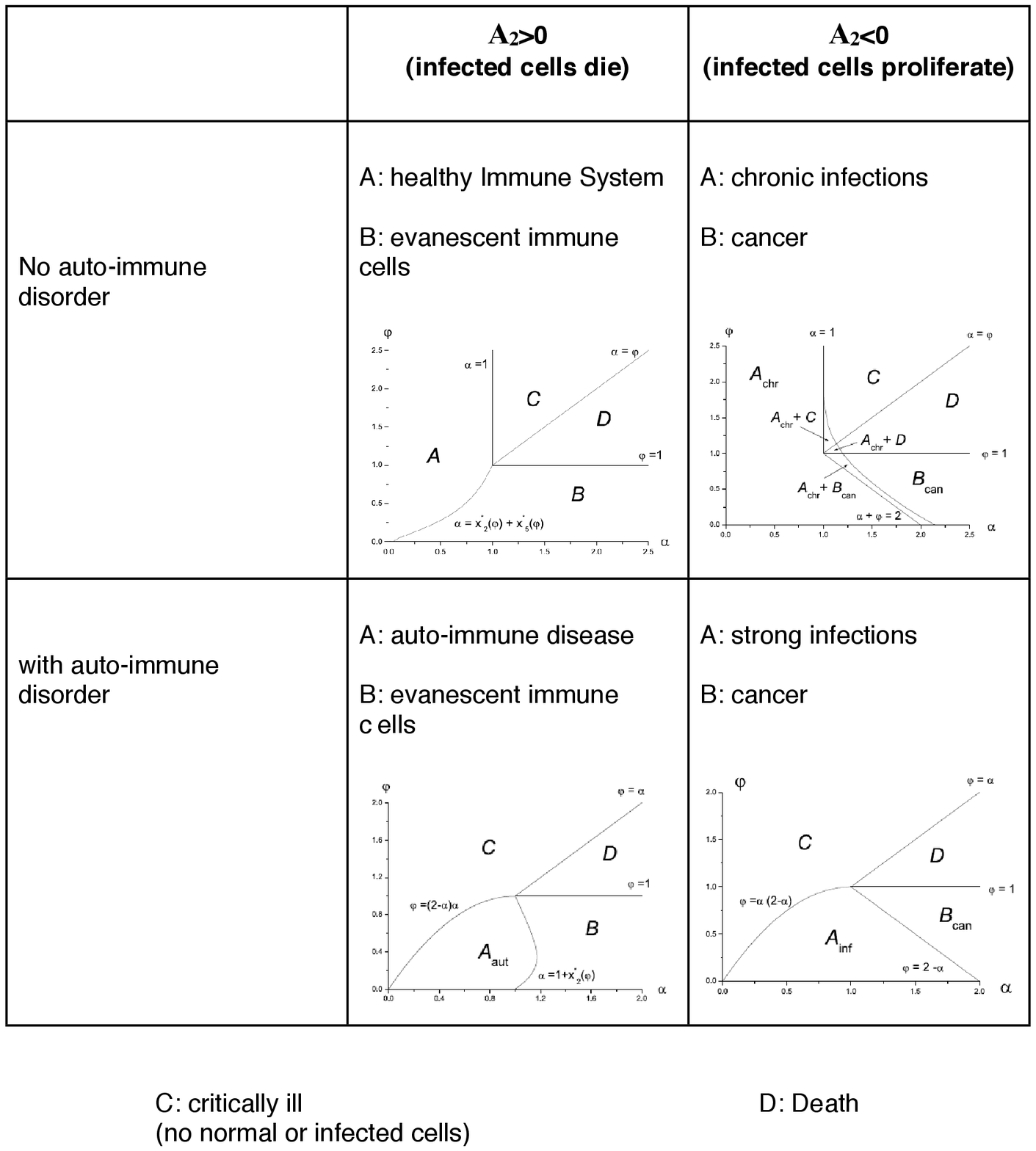}}
\caption{Synthetic table summary of the different states of the organism.
$C$ denotes the critically ill state. $D$ represents death.}
\label{fig:Fig.11}
\end{figure}

\end{document}